\documentclass{aa}
\usepackage{graphics}
\usepackage{psfig}
\newcommand{\sbs}{SBS 1415+437}

\def\h2{H{\small II}}
\begin{document}

\title{Spectroscopic and photometric studies of low-metallicity 
star-forming dwarf  galaxies. III. 
SBS 1415+437}
\author{N. G.\ Guseva \inst{1}
\and P.\ Papaderos \inst{2}
\and Y. I.\ Izotov \inst{1}
\and R. F. Green \inst{3}
\and K. J.\ Fricke   \inst{2}
\and T. X.\ Thuan\inst{4}
\and K. G.\ Noeske\inst{2}}
\offprints{N.G. Guseva, guseva@mao.kiev.ua}
\institute{      Main Astronomical Observatory,
                 Ukrainian National Academy of Sciences,
                 Zabolotnoho 27, Kyiv 03680,  Ukraine
\and
                 Universit\"ats--Sternwarte, Geismarlandstra\ss e 11,
                 D--37083 G\"ottingen, Germany
\and
                 National Optical Astronomy Observatory, 
                 Tucson, AZ 85726, USA
\and
                 Astronomy Department, University of Virginia, 
                 Charlottesville, VA 22903, USA
}

\date{Received \hskip 2cm; Accepted}
\titlerunning{A spectroscopic and photometric study of SBS 1415+437}
\authorrunning{N. G. Guseva et al.}

\abstract{We present a detailed optical spectroscopic and $B$, $V$, $I$, H$\alpha$ 
photometric study of the metal-deficient cometary blue compact dwarf (BCD) galaxy 
SBS 1415+437. We derive an oxygen abundance 12 + log(O/H) = 7.61$\pm$0.01 
and 7.62$\pm$0.03 ($Z$ = $Z_\odot$/20)\thanks{12+log(O/H)$_{\odot}$ = 8.92
(Anders \& Grevesse \cite{Anders89}).} in the two brightest H {\sc ii} 
regions, 
among the lowest in BCDs. The helium mass fractions in these regions
are $Y$ = 0.246$\pm$0.003 and 0.243$\pm$0.010.
Four techniques based on the equivalent widths of the hydrogen emission and 
absorption lines, the spectral energy distribution and 
the colours of the galaxy are used to put
constraints on the age of the stellar population in the low-surface-brightness
(LSB) component of the galaxy, 
assuming two limiting cases of star formation (SF), the
case of an instantaneous burst and that of a continuous SF with 
a constant or a variable star formation rate 
(SFR). The spectroscopic and photometric data for different regions of the LSB
component are well reproduced by a young stellar population with an
age $t$ $\leq$ 250 Myr, assuming a small  extinction in the range
$A_V$ = 0 -- 0.6 mag. 
Assuming no extinction, we find that the upper limit for the mass of the old
stellar population, formed between 2.5 Gyr and 10 Gyr, is 
not greater than $\sim$ (1/20 -- 1) of that of the stellar population formed 
during the last $\sim$ 250 Myr. Depending on the region considered, this also 
implies that the SFR in the most recent SF period 
must be 20 to 1000 times greater than the SFR at ages $\ga$ 2.5 Gyr.
We compare the photometric and spectroscopic properties of \sbs\ with
those of a sample of 26 low-metallicity dwarf irregular and BCD galaxies.
We show that there is a clear trend for 
the stellar LSB component of lower-metallicity galaxies to be 
 bluer. This trend cannot be explained only by   
metallicity effects. There must be also a change in the age of the
stellar populations. The most metal-deficient galaxies have also smaller
luminosity-weighted ages.
}

\maketitle

\markboth {N.G. Guseva et al.}{Spectroscopic and photometric studies of low-metallicity star-forming dwarf galaxies. III. SBS 1415+437}

\section {Introduction}
\label{intro}

Since its discovery as a metal-deficient
blue compact dwarf (BCD) galaxy (Thuan, Izotov \& Lipovetsky \cite{til95}),
SBS 1415+437 ($\equiv$ CG 389)
has been considered as a probable nearby young dwarf galaxy. Situated at a 
distance $D$ = 11.4 Mpc it was classified by Thuan, Izotov \& Foltz 
(\cite{ti99}) as a cometary BCD with a very bright supergiant H {\sc ii} region
at the SW tip of the galaxy.

From 
4m Kitt Peak National Observatory (KPNO) telescope spectra, Thuan et al. 
(\cite{til95}) first derived 
an oxygen abundance of 12 + log(O/H) = 7.51$\pm$0.01 in SBS 1415+437 placing the 
galaxy among the most metal-deficient BCDs known. Later, Izotov \& Thuan 
(\cite{IT98,IT99}) derived from the same spectrum
12 + log(O/H) = 7.59$\pm$0.01 
using five-level atom models for abundance determination
instead of the three-level atom model used by Thuan et al. (\cite{til95}).
Thuan et al. (\cite{ti99}), using 
Multiple Mirror Telescope (MMT) and {\sl Hubble
Space Telescope} ({\sl HST}) FOS observations, derived 12 + log(O/H) = 
7.60$\pm$0.01 and 7.54$\pm$0.14, respectively. The high brightness and 
low metallicity of the H {\sc ii} region in SBS 1415+437 make this galaxy
one of the best objects for helium abundance determination. Izotov \& Thuan
(\cite{IT98}) and Thuan et al. (\cite{ti99}) derived respectively a helium mass 
fraction $Y$ = 0.244$\pm$0.002 and 0.246$\pm$0.004 for it, close to
the primordial helium mass fraction of $Y_{\rm p}$ = 0.245$\pm$0.002 by 
Izotov et al. (\cite{ICFGGT99}).

Thuan et al. (\cite{ti99}) have discussed the evolutionary status of
SBS 1415+437, using ground-based MMT spectroscopic and {\sl HST}/WFPC2
photometric data. Based on the ($V-I$) vs $I$ colour-magnitude diagrams (CMD)
and spectral energy distributions (SED) in the optical range, they concluded
that SBS 1415+437 is a truly young galaxy that did not start to form stars
until $\sim$ 100 Myr ago. However, the $V$ and $I$ images used by Thuan et al.
(\cite{ti99}) were not deep enough for the detection of old red giant 
branch (RGB) stars in the CMD.
Furthermore, they considered an instantaneous burst model which gives 
only a lower limit to the age of the stellar population in SBS 1415+437.

In this paper we combine new spectroscopic and photometric data with
previous observations to derive elemental abundances and to better
constrain the age of the stellar population in SBS 1415+437. For the latter task
we use four different techniques of age determination and consider
different star formation (SF) histories.
The paper is organized as follows. In Sect. \ref{obs} we describe the 
observations and data reduction. The photometric properties of \sbs\ 
are described in Sect. \ref{phot}. We derive in Sect. \ref{chem} 
the chemical abundances 
in the two brightest H {\sc ii} regions. In Sect. \ref{age} we discuss the 
properties of the stellar populations in \sbs\ and compare them with those 
in other low-metallicity dwarf galaxies.
Finally, 
Sect. \ref{conc} summarises the main conclusions of this study.

\section{Observations and data reduction \label{obs}} 

\subsection{Photometry}

Narrow-band images of SBS 1415+437 in the 
H$\alpha$ line at $\lambda$6563\AA\ 
through a passband with a full width at half maximum
(FWHM)
of 74\AA, and 
in the adjacent continuum at $\lambda$6477\AA\ through a passband with 
FWHM = 72\AA\ were obtained with the Kitt Peak\footnote{Kitt Peak National 
Observatory (KPNO) is operated by the Association of Universities for Research
in Astronomy (AURA), Inc., under cooperative 
agreement with the National Science 
Foundation (NSF).} 2.1m telescope 
on April 22, 1999 during a photometric night. 
The telescope was equipped with a Tektronix 1024 $\times$ 1024
CCD detector operating at a gain of 3\,e$^-$\,ADU$^{-1}$, giving an 
instrumental scale of 0\farcs 305 pixel$^{-1}$ and field of view of 5\arcmin. 
The total exposures of 50 min in the H$\alpha$ line and 40 min 
in the  
adjacent continuum bluewards of H$\alpha$ were split up into 5 and 4 
subexposures, slightly offset with respect to each other for removal
of cosmic particle hits and bad pixels. The point spread function 
has a FWHM of 2\farcs2.
Bias and flat--field frames were obtained during the same night.
The standard stars Feige 34 and HZ 44 were observed
in both filters during the same night at several airmasses for 
absolute flux calibration.

Another broad-band $B$ image  (15 min) of 
SBS 1415+437 was obtained on March~9, 1997 under
photometric conditions, with the CAFOS focal reducer 
attached to the 2.2m telescope of the German-Spanish Astronomical 
Center, Calar Alto\footnote{German--Spanish Astronomical 
Center, Calar Alto, operated by the Max--Planck--Institute for Astronomy, 
Heidelberg, jointly with the Spanish National Commission for Astronomy.}, Spain. 
CAFOS was equipped with a SITe 2048 $\times$ 2048 CCD operating at a gain 
of 2.3 e$^{-}$ ADU$^{-1}$, with a read-out noise of $<$ 3 counts (rms). 
With a focal ratio of f/4.4, the instrumental scale was 0\farcs53 pixel$^{-1}$ 
and field of view  $\sim$~15\arcmin. The seeing during
the observations was 3\farcs1 (FWHM).
Standard stars from Christian et al. (1985) were observed at several
airmasses for calibration.

Standard reduction steps, including bias subtraction, flat--field 
correction, removal of cosmic particle hits 
and absolute flux calibration were carried out using  
IRAF\footnote{IRAF is the Image 
Reduction and Analysis Facility distributed by the 
National Optical Astronomy Observatory, which is operated by the 
AURA under 
cooperative agreement with the NSF.}
and MIDAS\footnote{Munich Image Data Analysis System, provided by the 
European Southern Observatory (ESO).}.

The ground-based photometric data  were supplemented by the 
{\sl HST}/WFPC2 $V$ (F569W) and $I$ (F791W) images
described by Thuan et al. (\cite{ti99}).

\begin{figure}
\begin{picture}(10.,10.)
\put(0,0){{\psfig{figure=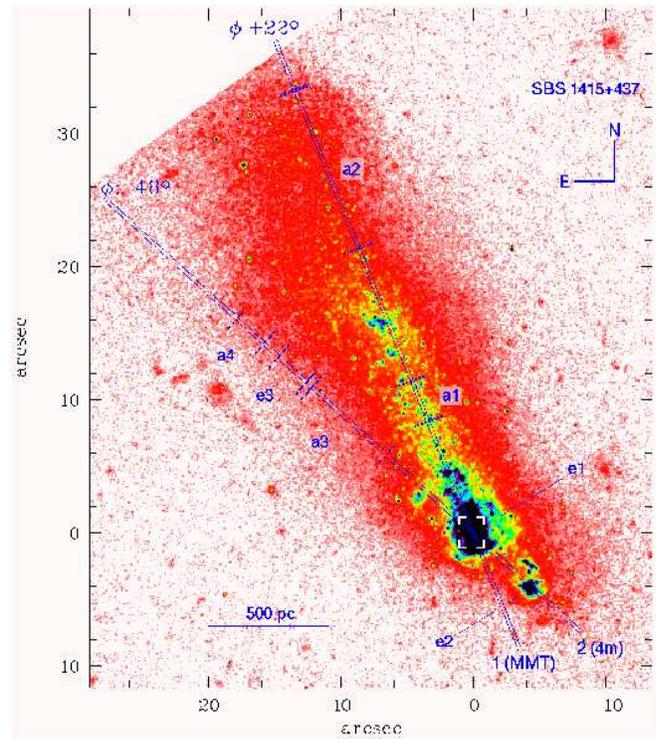,height=10.cm,angle=0,clip=}}}
\end{picture}
\caption[]{{\sl HST}~ $V$ image of SBS 1415+437. The long-slit positions 
during the two observations with the MMT and
4m Mayall telescope are labeled ``1'' and ``2'' respectively. North is up and east 
is to the left. Regions {\it e}1 -- {\it e}3 with hydrogen emission lines
in the spectra and regions {\it a}1 -- {\it a}4 with hydrogen absorption lines
are labeled.} 
\label{f1}
\end{figure}

\begin{figure}
\begin{picture}(10.,11.)
\put(0,0){{\psfig{figure=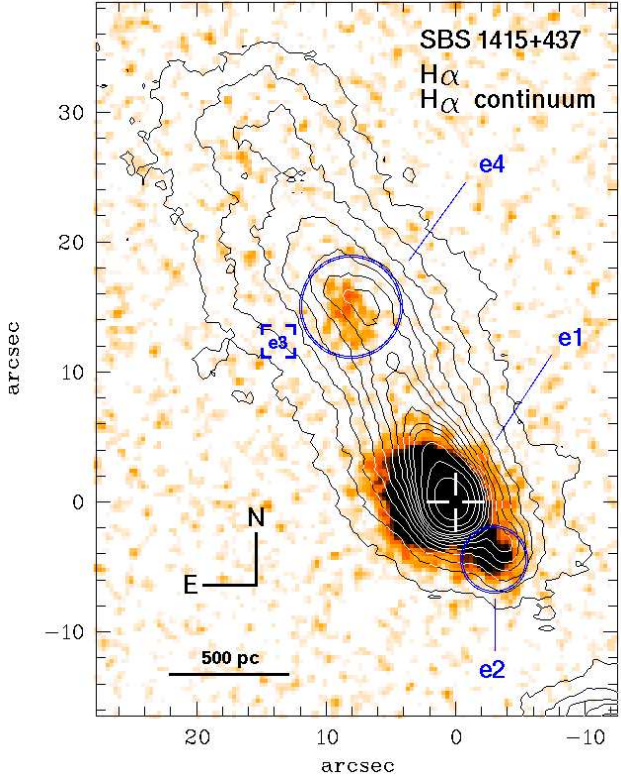,height=10.5cm,angle=0,clip=}}}
\end{picture}
\caption[]{Continuum-subtracted H$\alpha$ image of SBS 1415+437 with superposed 
 H$\alpha$ continuum isophotes. Regions with nebular emission are 
labeled {\it e}1, {\it e}2
and {\it e}4. The faint region {\it e}3 with H$\beta$ and H$\alpha$
emission lines in its spectrum is not seen in the H$\alpha$ image.} 
\label{halpha}
\end{figure}

\begin{figure}
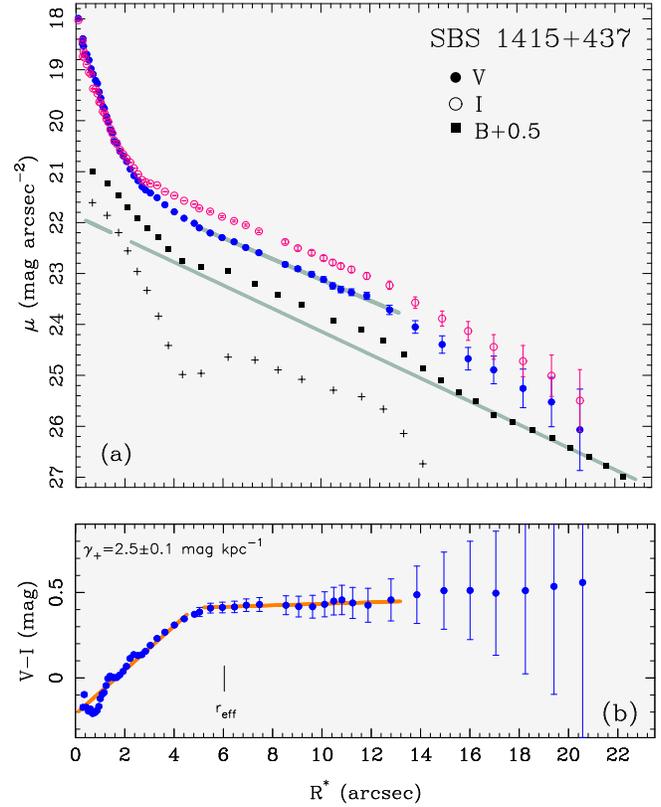

\begin{picture}(10.,11.2)
\put(0,4.2){{\psfig{figure=3331.f3a.ps,width=8.5cm,angle=270,clip=}}}
\put(0.04,0.){{\psfig{figure=3331.f3b.ps,width=8.5cm,angle=270,clip=}}}
\end{picture}
\caption[]{({\bf a}) 
Surface brightness profiles (SBPs) of SBS 1415+437 in
$V$ and $I$ (filled and open circles, respectively), derived from 
{\sl HST}/WFPC2 data. A linear fit to the $V$ SBP in the radius interval
4\arcsec$\leq R^*\la$13\arcsec\ is shown by the upper solid-grey line.
Both SBPs reveal a significant slope change for $R^*\ga$13\arcsec.
Filled squares show the ground-based $B$ SBP,
shifted vertically by 0.5 mag. In addition to the exponential 
intensity slope at radii 4\arcsec$\leq R^*\la$13\arcsec, this profile
reveals for $R^*\ga$16\arcsec\ an outer exponential regime with an 
$\alpha\sim$0.27 kpc (lower solid-grey line).
The residuals after subtraction of the outer exponential distribution 
from the $B$ SBP 
are shown by crosses.
({\bf b}) ($V-I$) colour profile of SBS 1415+437, computed from subtraction 
of the SBPs in ({\bf a}). A strong colour gradient
 $\gamma_+$ = 2.5 mag kpc$^{-1}$ is derived for radii smaller than the  
effective radius $r_{\rm eff}$ (inner fit). The ($V-I$) 
colour of 0.4 -- 0.5 mag for larger radii is nearly constant 
with a $\gamma_+<$0.2 mag kpc$^{-1}$.
Comparison of the $B$ and $V$ SBPs implies a $B-V$ color $\sim$0.2 mag
in the radius range $13\arcsec \leq R^* \leq 16\arcsec$.
}
\label{sbp}
\end{figure}

\subsection{Spectroscopy}

Spectroscopic observations were carried out on June 18, 1999, at the 
Kitt Peak 4m 
Mayall telescope with the Ritchey-Chr\'etien spectrograph and a
T2KB 2048~$\times$~2048 CCD detector. The 2\arcsec~$\times$~300\arcsec\
slit was centered on the brightest H {\sc ii} region {\it e}1 
(slit 2 in Fig. \ref{f1}) with position angle P.A. = 48$^{\circ}$ 
so as to include the second brightest H {\sc ii} region 
{\it e}2 to the SW of region {\it e}1. 
We used the KPC-10A grating in first order and a GG 375 order separation filter.
The spatial scale along the slit
was 0\farcs69 pixel$^{-1}$ and the spectral resolution $\sim$7~\AA\ (FWHM).
The spectra were obtained at an airmass 1.27. The total 
exposure time of 60 minutes was broken up into 3 subexposures. 
No correction for atmospheric refraction was made because of 
the small airmass during the observations.
Two Kitt Peak 
spectrophotometric standard stars were observed for flux calibration.
For wavelength calibration, He-Ne-Ar comparison spectra were obtained
after each exposure.

\begin{figure*}
   \hspace*{2.5cm}\psfig{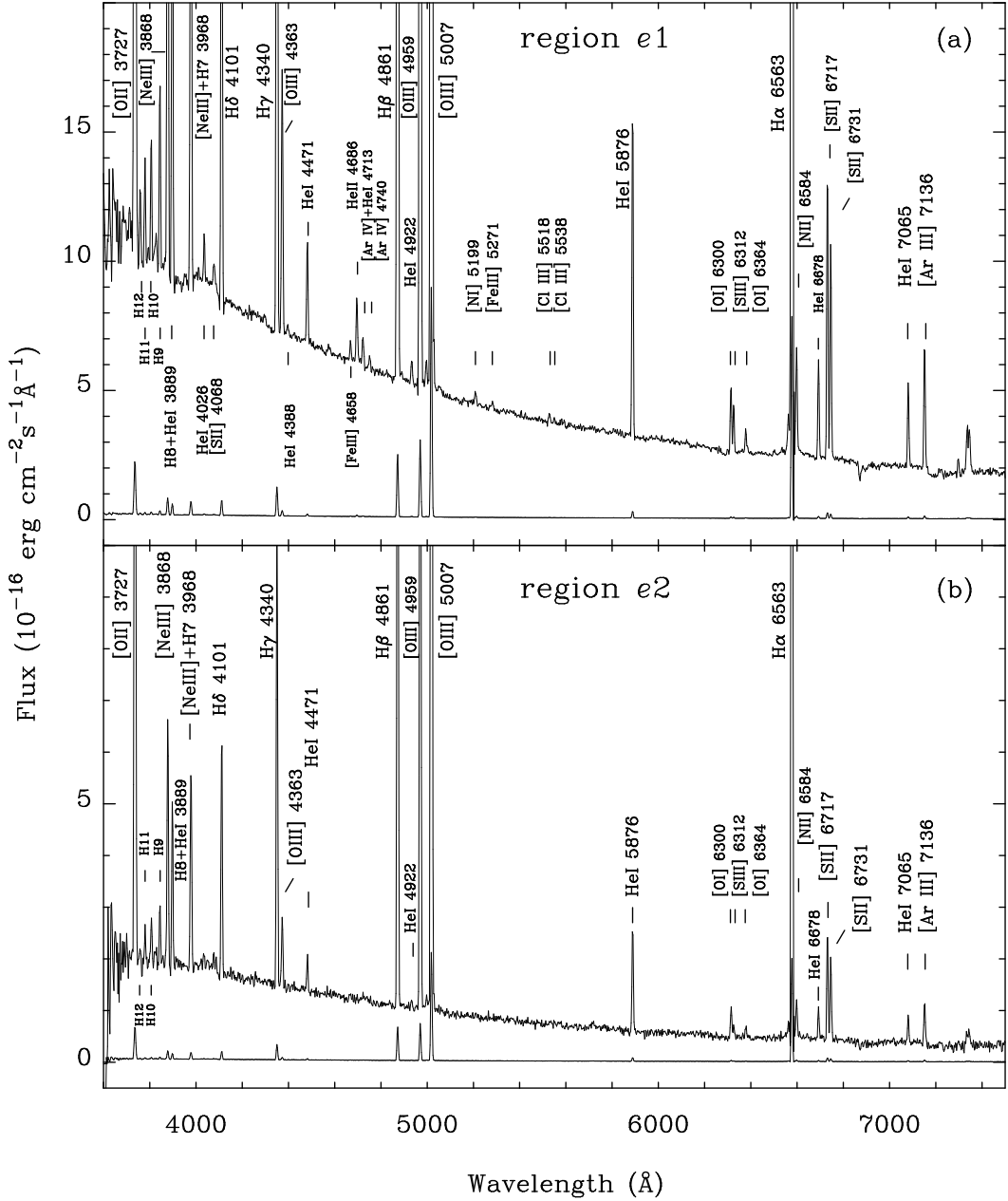}
    \caption{The KPNO 4m telescope spectra of the brightest H {\sc ii} 
regions {\it e}1 and {\it e}2 with the identified emission lines. 
The lower spectra in ({\bf a}) and ({\bf b}) are the 
observed spectra downscaled by factors of 50 and 30, respectively.
      }
    \label{fig:brightsp}
\end{figure*}

The data reduction was performed with the IRAF
software package. This includes  bias--subtraction, 
flat--field correction, cosmic-ray removal, wavelength calibration, 
night sky background subtraction, correction for atmospheric extinction and 
absolute flux calibration of the two--dimensional spectrum.

For abundance determination, one-dimensional spectra of regions 
{\it e}1 and {\it e}2 
were extracted within 
apertures of 2\arcsec\ $\times$ 4\farcs6 and 2\arcsec\ $\times$ 4\farcs0, 
respectively. 
In addition, we extracted spectra of the low-surface-brightness (LSB)
 regions {\it a}3 and {\it a}4 with strong hydrogen 
Balmer absorption lines,
and of region {\it e}3 with H$\alpha$ and H$\beta$ 
in emission (Fig. \ref{f1}).

We also used the two-dimensional MMT spectrum obtained by Thuan et al. (\cite{ti99})
with the slit oriented at P.A. = 22$^\circ$ (slit 1 in Fig. \ref{f1}).
We extracted one-dimensional spectra of the LSB regions {\it a}1 and 
{\it a}2 within apertures of 1\farcs5 $\times$ 3\farcs4 and 
1\farcs5 $\times$ 13\farcs2, respectively (Fig. \ref{f1}). 
These spectra show strong hydrogen Balmer absorption lines.

The selected LSB regions 
are listed in Tables \ref{t:emhahb} and \ref{t:abshdhg} 
with their positions and spatial extents. Origins are set at the 
center of the brightest region {\it e}1 (Fig.~\ref{f1},~\ref{halpha}).
The spectra of the LSB regions are used to study the stellar populations
and to constrain the age of the oldest stars which contribute to the 
light of these regions.

\section{Photometric properties \label{phot}}

It is seen from the continuum-subtracted H$\alpha$ images (Fig. 2)
that star-forming activity in SBS 1415+437 is primarily occurring 
in regions {\it e}1 and {\it e}2, with some additional faint H$\alpha$ 
emission present in region {\it e}4. 
However, the available narrow-band data are not deep enough for tracing
faint H$\alpha$ emission in other regions of the LSB component, such as in
regions {\it a}1,~{\it a}3 and {\it e}3 where H$\alpha$ and H$\beta$ have 
been detected spectroscopically (Table~\ref{t:emhahb}).
 The latter fact suggests that some low-level SF may be present at 
various locations within the LSB component.

The photometric properties of the stellar LSB component of SBS 1415+437 
were first investigated by Thuan et al. (1999) using
{\sl HST}/WFPC2 $V$ and $I$ images. 
These authors fit the surface brightness profiles (SBPs) 
of SBS 1415+437 with an exponential distribution 
in the radius range 4\arcsec$\leq R^*\la$13\arcsec\ 
with a scale length $\alpha$ = 5\farcs 4 ($\approx$0.3 kpc). 
However, their study was limited by the fact 
that the {\sl HST} images do not include the outermost NE part of the LSB 
component. 
Furthermore, the SBPs in Thuan et al. (1999) reach only 
a surface brightness level $\mu\sim 24.5$ mag arcsec$^{-2}$. 
It is known, however, that the star-forming component may contribute 
to the optical BCD emission to fainter surface brightness levels
(see e.g. Papaderos et al. 2002 and references therein). 
A comparatively young stellar population has 
been observed in the inner part of the LSB component of other cometary BCDs, 
several hundred pc away from the brightest \ion{H}{ii} region 
(Noeske et al. 2000; Guseva et al. 2001; Fricke et al. 2001).
Therefore deeper images are needed to study the outer parts of 
the LSB component in \sbs.

 Note that the SBPs by Thuan et al. (1999) show 
in the outermost part ($13\arcsec \la R^*\la 16$\arcsec, or 
$24 \la \mu_V \la 24.5$) a steeper 
exponential intensity decrease than the one observed at intermediate intensity levels.
This slope change, not discussed in Thuan et al. (1999), is found
independently by us in the {\sl HST}/WFPC2 SBPs derived with 
the method~iv of Papaderos et al. (2002) and
ellipse fitting to the visible
part of the LSB component. While 
the latter method extends surface photometry out to 
larger radii ($R^*\sim$ 22\arcsec), it is subject to large uncertainties
because the NE part of the LSB component with 
$\mu$ $\ga$24.5 $V$ mag arcsec$^{-2}$ ($R^*\ga$ 16\arcsec)
lies outside the {\sl HST}/WFPC2 
field of view. 
 Additionally, the outermost LSB isophotes show considerable 
deviations from ellipticity.
 
 To study the surface brightness distribution at large radii we 
use the ground-based Calar Alto $B$ image. Despite the poor 
spatial resolution this image allows us to study the 
entire LSB component out to its Holmberg radius.
The change in the exponential slope for
$13\arcsec \la R^*\la 16 \arcsec $ is confirmed from the ground-based $B$ SBP.
At large radii, however, this SBP reveals a flatter, outer exponential 
regime with a scale length $\alpha$ fairly comparable to that previously 
obtained at intermediate intensity levels from {\sl HST} data 
(upper thick-grey line in Fig. 3a). 
From fitting an exponential model to the $B$ SBP for $R^*\geq$16\arcsec\
(lower thick-grey line in Fig. 3a) we obtain a central surface brightness 
$\mu_{\rm E,0}=21.37$ $B$ mag arcsec$^{-2}$ and a scale length $\alpha=0.27$
kpc.

Note, however, that the inner exponential profile studied by Thuan et
al. (1999) is $\ga$0.3 mag brighter than the outer one, which suggests that 
more than 1/4 of the emission associated with this profile originates 
from the part of the LSB component
between regions {\it e}1 and {\it a}2.

The present data provide no compelling evidence for a large age difference
between the stellar population which dominates within the inner exponential regime
discussed in Thuan et al. (1999) and that responsible for the outermost LSB
emission (i.e. for $R^*>16$\arcsec).
The ($V-I$) profile reveals a strong colour 
gradient ($\gamma_+$=2.5 mag kpc$^{-1}$; inner solid-grey
line in Fig.~\ref{sbp}b) within the inner 5\arcsec, or 
roughly the $V$ band effective radius 
($r_{\rm eff}=5\farcs 6$) of \sbs. At larger radii, however, 
linear fits to the ($V-I$) profile yield, depending on whether they 
are error-weighted or not, a gradient not exceeding 0.1 and 0.2 
mag kpc$^{-1}$, respectively. Comparable values are also found from subtraction 
of the exponential fits in $V$ and $I$ (Fig.~\ref{sbp}a) in the 
radius range $r_{\rm eff}\leq R^* \leq 13\arcsec$ (Fig.~\ref{sbp}b, 
solid-grey line at 
intermediate radii). 

\section{Chemical abundances \label{chem}}

 In this section we derive the elemental abundances of
regions {\it e}1 and {\it e}2 using the Kitt Peak 4m telescope observations.
Their spectra with strong emission lines are shown in Fig. \ref{fig:brightsp}.


\begin{table*}[tbh]
\caption{Observed ($F$($\lambda$)) and extinction-corrected  
($I$($\lambda$)) fluxes and equivalent widths ($EW$) of emission lines
 in the H {\sc ii} regions {\it e}1 and {\it e}2.}
\label{t:Intens}
\begin{tabular}{lccrcccr} \hline \hline
  &\multicolumn{3}{c}{region {\it e}1}&&\multicolumn{3}{c}{region {\it e}2} \\ \cline{2-4} \cline{6-8}
$\lambda_{0}$(\AA) Ion                  &$F$($\lambda$)/$F$(H$\beta$)
&$I$($\lambda$)/$I$(H$\beta$)&$EW$(\AA)&&$F$($\lambda$)/$F$(H$\beta$)&$I$($\lambda$)/$I$(H$\beta$) 
&$EW$(\AA)  \\ \hline
3727\ [O {\sc ii}]                & 0.968 $\pm$0.004 & 0.949 $\pm$0.004 &88.2 $\pm$0.3 && 1.099 $\pm$0.016 &1.085 $\pm$0.016 &79.2 $\pm$0.8 \\
3750\ H12                         & 0.024 $\pm$0.001 & 0.060 $\pm$0.004 & 2.2 $\pm$0.1 && 0.007 $\pm$0.004 &0.031 $\pm$0.021 &0.5  $\pm$0.3 \\
3771\ H11                         & 0.029 $\pm$0.001 & 0.064 $\pm$0.003 & 2.7 $\pm$0.1 && 0.030 $\pm$0.004 &0.053 $\pm$0.009 &2.3  $\pm$0.3 \\
3798\ H10                         & 0.038 $\pm$0.001 & 0.073 $\pm$0.003 & 3.5 $\pm$0.1 && 0.045 $\pm$0.005 &0.067 $\pm$0.009 &3.4  $\pm$0.4 \\
3835\ H9                          & 0.057 $\pm$0.001 & 0.091 $\pm$0.003 & 5.3 $\pm$0.1 && 0.055 $\pm$0.005 &0.078 $\pm$0.008 &4.2  $\pm$0.4 \\
3868\ [Ne {\sc iii}]              & 0.262 $\pm$0.002 & 0.257 $\pm$0.002 &25.2 $\pm$0.2 && 0.241 $\pm$0.007 &0.238 $\pm$0.007&18.4 $\pm$0.5  \\
3889\ H8\ +\ He {\sc i}           & 0.166 $\pm$0.002 & 0.196 $\pm$0.002 &16.4 $\pm$0.2 && 0.151 $\pm$0.006 &0.172 $\pm$0.008&11.6 $\pm$0.4  \\
3968\ [Ne {\sc iii}]\ +\ H7       & 0.227 $\pm$0.002 & 0.255 $\pm$0.003 &23.0 $\pm$0.2 && 0.217 $\pm$0.007 &0.236 $\pm$0.008&17.2 $\pm$0.5  \\
4026\ He {\sc i}                  & 0.013 $\pm$0.001 & 0.013 $\pm$0.001 & 1.3 $\pm$0.1 &&       ...        &       ...       &\multicolumn {1}{c}{...}\\
4068\ [S {\sc ii}]                & 0.008 $\pm$0.001 & 0.008 $\pm$0.001 & 0.8 $\pm$0.1 &&       ...        &       ...       &\multicolumn {1}{c}{...}\\
4101\ H$\delta$                   & 0.245 $\pm$0.002 & 0.269 $\pm$0.002 &27.6 $\pm$0.2 && 0.236 $\pm$0.007 &0.253 $\pm$0.008&19.9 $\pm$0.5  \\
4340\ H$\gamma$                   & 0.452 $\pm$0.002 & 0.469 $\pm$0.003 &56.9 $\pm$0.3 && 0.446 $\pm$0.009 &0.458 $\pm$0.010&42.9 $\pm$0.7  \\
4363\ [O {\sc iii}]               & 0.085 $\pm$0.001 & 0.084 $\pm$0.001 &10.8 $\pm$0.2 && 0.071 $\pm$0.005 &0.070 $\pm$0.005 &6.9 $\pm$0.5  \\
4388\ He {\sc i}                  & 0.004 $\pm$0.001 & 0.004 $\pm$0.001 & 0.5 $\pm$0.2 &&       ...        &       ...       &\multicolumn {1}{c}{...}\\
4471\ He {\sc i}                  & 0.034 $\pm$0.001 & 0.033 $\pm$0.001 & 4.5 $\pm$0.2 && 0.034 $\pm$0.004 &0.034 $\pm$0.004 &3.6 $\pm$0.4  \\
4658\ [Fe {\sc iii}]              & 0.006 $\pm$0.001 & 0.006 $\pm$0.001 & 1.0 $\pm$0.1 &&       ...        &       ...       &\multicolumn {1}{c}{...}\\
4686\ He {\sc ii}                 & 0.023 $\pm$0.001 & 0.023 $\pm$0.001 & 3.5 $\pm$0.2 &&       ...        &       ...       &\multicolumn {1}{c}{...}\\
4713\ [Ar {\sc iv}]\ +\ He {\sc i}& 0.011 $\pm$0.001 & 0.011 $\pm$0.001 & 1.7 $\pm$0.2 &&       ...        &       ...       &\multicolumn {1}{c}{...}\\
4740\ [Ar {\sc iv}]               & 0.005 $\pm$0.001 & 0.005 $\pm$0.001 & 0.8 $\pm$0.2 &&       ...        &       ...       &\multicolumn {1}{c}{...}\\
4861\ H$\beta$                    & 1.000 $\pm$0.004 & 1.000 $\pm$0.004 &166.2 $\pm$0.5&& 1.000 $\pm$0.014 &1.000 $\pm$0.014&133.8 $\pm$1.3 \\
4922\ He {\sc i}                  & 0.008 $\pm$0.001 & 0.008 $\pm$0.001 & 1.4 $\pm$0.2 && 0.012 $\pm$0.004 &0.012 $\pm$0.004 &1.7  $\pm$0.5 \\
4959\ [O {\sc iii}]               & 1.203 $\pm$0.005 & 1.179 $\pm$0.005 &214.2 $\pm$0.6&& 1.082 $\pm$0.015 &1.068 $\pm$0.015&156.0 $\pm$1.5 \\
5007\ [O {\sc iii}]               & 3.613 $\pm$0.012 & 3.542 $\pm$0.012 &663.0 $\pm$0.9&&3.210 $\pm$0.036 &3.169 $\pm$0.036 &473.6 $\pm$2.4 \\
5199\ [N {\sc i}]                 & 0.005 $\pm$0.001 & 0.005 $\pm$0.001 &1.0 $\pm$0.2  &&       ...        &      ...        &\multicolumn {1}{c}{...}\\
5271\ [Fe {\sc iii}]              & 0.003 $\pm$0.001 & 0.003 $\pm$0.001 & 0.7 $\pm$0.2 &&       ...        &      ...        &\multicolumn {1}{c}{...}\\
5518\ [Cl {\sc iii}]              & 0.003 $\pm$0.001 & 0.003 $\pm$0.001 &  0.8 $\pm$0.2&&       ...        &      ...        &\multicolumn {1}{c}{...}\\
5538\ [Cl {\sc iii}]              & 0.002 $\pm$0.001 & 0.002 $\pm$0.001 &  0.4 $\pm$0.2&&       ...        &      ...        &\multicolumn {1}{c}{...}\\
5876\ He {\sc i}                  & 0.100 $\pm$0.001 & 0.098 $\pm$0.001 &28.8 $\pm$0.3 && 0.107 $\pm$0.004 &0.106 $\pm$0.004 &26.2 $\pm$1.0 \\
6300\ [O {\sc i}]                 & 0.022 $\pm$0.001 & 0.022 $\pm$0.001 & 8.0 $\pm$0.3 && 0.032 $\pm$0.003 &0.032 $\pm$0.003 &9.8  $\pm$0.9 \\
6312\ [S {\sc iii}]               & 0.015 $\pm$0.001 & 0.015 $\pm$0.001 & 5.3 $\pm$0.2 && 0.012 $\pm$0.002 &0.011 $\pm$0.002 &3.5  $\pm$0.7 \\
6364\ [O {\sc i}]                 & 0.008 $\pm$0.001 & 0.008 $\pm$0.001 & 2.7 $\pm$0.2 && 0.011 $\pm$0.002 &0.011 $\pm$0.002 &3.4  $\pm$0.7 \\
6563\ H$\alpha$                   & 2.714 $\pm$0.009 &2.670 $\pm$0.010 &997.9 $\pm$1.7 &&2.810 $\pm$0.031 &2.780 $\pm$0.034 &871.7 $\pm$4.5 \\
6584\ [N {\sc ii}]                & 0.033 $\pm$0.001 & 0.032 $\pm$0.001 &11.1 $\pm$0.3 && 0.042 $\pm$0.003 &0.042 $\pm$0.003 &13.2 $\pm$1.0 \\
6678\ He {\sc i}                  & 0.029 $\pm$0.001 & 0.028 $\pm$0.001 &11.1 $\pm$0.3 && 0.029 $\pm$0.003 &0.028 $\pm$0.003 & 9.2 $\pm$0.8 \\
6717\ [S {\sc ii}]                & 0.090 $\pm$0.001 & 0.088 $\pm$0.001 &34.1 $\pm$0.4 && 0.107 $\pm$0.004 &0.105 $\pm$0.004 &34.3 $\pm$1.2 \\
6731\ [S {\sc ii}]                & 0.066 $\pm$0.001 & 0.065 $\pm$0.001 &25.2 $\pm$0.4 && 0.080 $\pm$0.004 &0.079 $\pm$0.004 &25.9 $\pm$1.1 \\
7065\ He {\sc i}                  & 0.027 $\pm$0.001 & 0.026 $\pm$0.001 &11.8 $\pm$0.3 && 0.027 $\pm$0.002 &0.027 $\pm$0.002 &8.4  $\pm$0.9 \\
7136\ [Ar {\sc iii}]              & 0.043 $\pm$0.001 & 0.042 $\pm$0.001 &20.6 $\pm$0.4 && 0.048 $\pm$0.003 &0.048 $\pm$0.003 &21.0 $\pm$1.3 \\
                     & & & & & & & \\
$C$(H$\beta$)\ dex             &\multicolumn {3}{c}{0.000$\pm$0.004} &&\multicolumn {3}{c}{0.000$\pm$0.014} \\
$F$(H$\beta$)$^{\rm a}$        &\multicolumn {3}{c}{4.54$\pm$0.01}   &&\multicolumn {3}{c}{0.71$\pm$0.01 } \\
$EW$(abs)~\AA                  &\multicolumn {3}{c}{3.4$\pm$0.1}  &&\multicolumn {3}{c}{1.8$\pm$0.3} \\
\hline
\end{tabular}

$^{\rm a}$in units 10$^{-14}$\ erg\ s$^{-1}$cm$^{-2}$.
\end{table*}


The observed ($F$($\lambda$)) and extinction-corrected 
($I$($\lambda$)) emission line fluxes relative to the H$\beta$ emission line 
fluxes, their equivalent widths $EW$, the extinction coefficients 
$C$(H$\beta$), the observed fluxes of the H$\beta$ emission line, and the 
equivalent widths of the hydrogen absorption lines
for regions {\it e}1 and {\it e}2 are shown in Table \ref{t:Intens}.
Despite the differences in aperture 
(2\arcsec\ $\times$ 4\farcs6 for the Kitt Peak 4m data,  1\farcs5\ $\times$ 5\arcsec\ and
1\farcs5\ $\times$ 0\farcs6\ for the MMT data from Thuan et al. \cite{ti99}), 
the relative fluxes of the emission lines for region {\it e}1
are in agreement within the errors with those derived by Thuan et al. (\cite{ti99}).

\begin{figure}[hbtp]
    \psfig{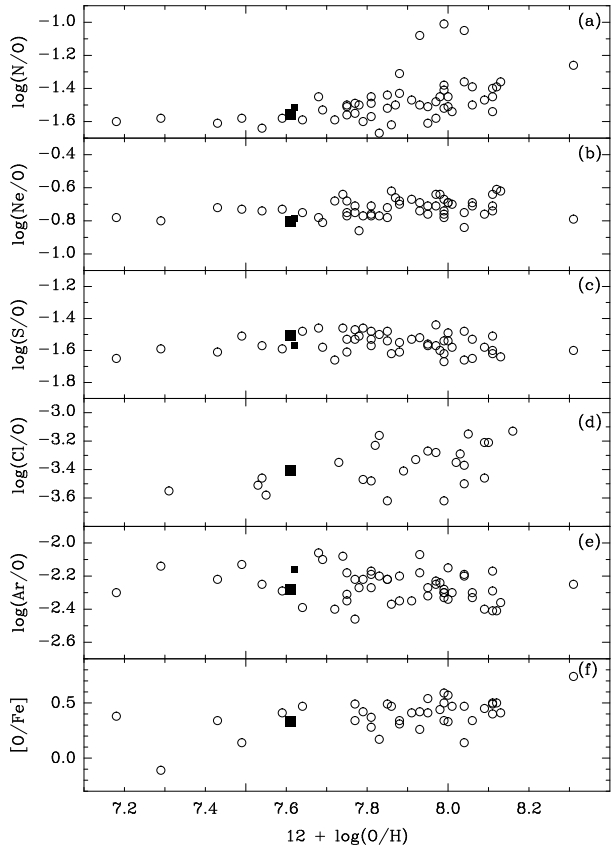}
    \caption{Comparison of the elemental abundance ratios, obtained for the brightest 
H {\sc ii} regions {\it e}1 (large squares) and {\it e}2 (small squares) 
with data for other BCDs (open circles).
      }
    \label{fig:heavy}
\end{figure}

The physical conditions and heavy element abundances in regions {\it e}1
and {\it e}2 were derived
following Izotov et al. (\cite{ITL94,ITL97a}) and Thuan et al. (\cite{til95}).
The electron temperatures $T_{\rm e}$(O {\sc iii}), $T_{\rm e}$(S {\sc iii}),
$T_{\rm e}$(O {\sc ii}) for the high-, intermediate- and low-ionization 
regions respectively, the electron number 
densities $N_{\rm e}$(S {\sc ii}), ionization correction factors (ICF), and
ionic and total heavy element abundances are shown in Table \ref{t:Chem} for 
both regions.
The oxygen abundance 12 + log(O/H) = 7.61 $\pm$ 0.01 ($Z_\odot$/20)
and heavy element abundance ratios for region {\it e}1 are in good agreement
with those derived by 
Thuan et al. (\cite{ti99}). The oxygen abundance 12 + log(O/H) =
7.62 $\pm$ 0.03 and heavy element abundance ratios in region {\it e}2 
are consistent with those for region {\it e}1 within the errors.

\begin{figure}[hbtp]
    \psfig{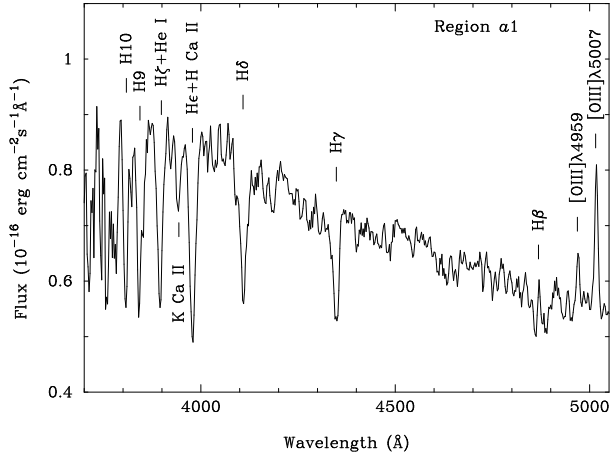}
    \caption{The blue part of the MMT spectrum of region {\it a}1 with 
labeled emission and absorption lines.
      }
    \label{fig:abs}
\end{figure}

 In Figure~\ref{fig:heavy} we compare the heavy element abundance 
ratios in the two brightest regions of 
\sbs\ with data for a sample of low-metallicity BCDs. 
The  Ne/O, S/O, Ar/O and [O/Fe] abundance ratios for the comparison sample are taken from 
Izotov $\&$ Thuan (1999), while the Cl/O abundance ratios are collected from
Izotov $\&$ Thuan (1998) and Izotov et al. (1997a).
 The heavy element abundance ratios for regions {\it e}1 (large squares)
and {\it e}2 (small squares) are in good agreement with those for other BCDs.
  Note, that the Cl/O ratio does not show any significant increase with increasing oxygen 
abundance. This conclusion is strengthened by the observations of 
Esteban et al. (1998, 1999a, 1999b) who derived  
log(Cl/O) in the range from --3.28 to --3.47 for 
high-metallicity H {\sc ii} regions  
in Orion, M17 and M8 with  12+log(O/H) =
8.60, 8.50 and 8.60, respectively. For comparison,  
log(Cl/O) = --3.37 is derived for the Sun (Anders $\&$ Grevesse \cite{Anders89}).

The high brightness of regions {\it e}1 and {\it e}2 
allows for a reliable determination of the $^4$He abundance. Nine He {\sc i}
emission lines are detected in the spectrum of region {\it e}1 (Table
\ref{t:Intens}). Two
of them, He {\sc i} $\lambda$3889 and $\lambda$4713, are blended with other
lines. Six He {\sc i} lines are detected in region {\it e}2.
The five brightest He {\sc i} $\lambda$3889, $\lambda$4471, $\lambda$5876, 
$\lambda$6678, $\lambda$7065 emission lines are used to correct 
their fluxes for collisional and fluorescent enhancement. This is
done by minimizing the deviations of the corrected He {\sc i} line
flux ratios from the recombination ratios, through varying the electron number
density in the He$^+$ zone and the optical depth in the He {\sc i} $\lambda$3889
emission line. The flux of this line was preliminarily corrected for the
contribution of the H {\sc i} $\lambda$3889 emission line, according to
prescriptions of Izotov et al. (\cite{ITL94,ITL97a}).
Helium abundances He$^+$/H$^+$, derived from the corrected He {\sc i} 
$\lambda$4471, $\lambda$5876, $\lambda$6678 line fluxes and their weighted
mean are listed in Table \ref{t:Chem}. The abundance He$^{+2}$/H$^+$ is
added to He$^{+}$/H$^+$ for region {\it e}1, as He {\sc ii} $\lambda$4686 is present in its
spectrum.  Note the lower He abundance derived from the He {\sc i} 
$\lambda$4471 flux which is most likely due to significant 
underlying stellar He {\sc i} $\lambda$4471 absorption.
The effect of underlying absorption for the other He {\sc i} emission
lines used in the He abundance determination is much smaller, as they
have much larger equivalent widths compared to the He {\sc i} 
$\lambda$4471 emission line (Table \ref{t:Intens}). 
The mean $^4$He mass
fractions $Y$ = 0.246$\pm$0.003 and 0.243$\pm$0.010 in regions {\it e}1 and 
{\it e}2 (Table \ref{t:Chem}) are
consistent with the values derived for \sbs\ by Izotov \& Thuan (\cite{IT98})
and Thuan et al. (\cite{ti99}). They are also consistent with the primordial 
$^4$He mass fraction $Y_{\rm p}$ = 0.244 $\pm$ 0.002, derived by extrapolating the 
$Y$ vs O/H linear regression to O/H = 0 (Izotov \& Thuan \cite{IT98}), or to 
$Y_{\rm p}$ = 0.245 $\pm$ 0.002 derived from spectroscopic observations of 
the two most metal-deficient BCDs known, 
I Zw 18 and SBS 0335--052 (Izotov et al. \cite{ICFGGT99}). 


\begin{table}[tbh]
\caption{Element abundances in regions {\it e}1 and {\it e}2.}
\label{t:Chem}
\begin{tabular}{lccc} \hline \hline
Value                               & region {\it e}1      && region {\it e}2  \\ \hline
$T_{\rm e}$(O {\sc iii})(K)               &16490$\pm$140   && 15910$\pm$540 \\
$T_{\rm e}$(O {\sc ii})(K)                &14430$\pm$110   && 14170$\pm$460 \\
$T_{\rm e}$(S {\sc iii})(K)               &15380$\pm$110   && 14900$\pm$450 \\
$N_{\rm e}$(S {\sc ii})(cm$^{-3}$)        & 60$\pm$30  &&  90$\pm$80   \\ \\
O$^+$/H$^+$($\times$10$^5$)         &0.930$\pm$0.020 && 1.127$\pm$0.101\\
O$^{+2}$/H$^+$($\times$10$^5$)      &3.071$\pm$0.063 && 2.997$\pm$0.256\\
O$^{+3}$/H$^+$($\times$10$^6$)      &0.994$\pm$0.057 &&      ...      \\
O/H($\times$10$^5$)                 &4.100$\pm$0.067 && 4.123$\pm$0.275\\
12 + log(O/H)                       &7.61$\pm$0.01   && 7.62$\pm$0.03\\ \\
N$^{+}$/H$^+$($\times$10$^7$)       &2.564$\pm$0.056  && 3.460$\pm$0.311\\
ICF(N)$^{\rm a}$                          &4.41~~~~~~~~~~~~&& 3.66~~~~~~~~~~~~\\
log(N/O)                             &--1.560$\pm$0.017~~&&--1.513$\pm$0.069~~\\ \\
Ne$^{+2}$/H$^+$($\times$10$^5$)     &0.480$\pm$0.011&& 0.491$\pm$0.045\\
ICF(Ne)$^{\rm a}$                         &1.34~~~~~~~~~~~~&&1.38~~~~~~~~~~~~\\
log(Ne/O)                            &--0.806$\pm$0.014~~&&--0.786$\pm$0.057~~\\ \\
S$^+$/H$^+$($\times$10$^7$)         &1.618$\pm$0.025  && 2.028$\pm$0.118\\
S$^{+2}$/H$^+$($\times$10$^7$)      &6.980$\pm$0.350  && 5.942$\pm$1.335\\
ICF(S)$^{\rm a}$                          &1.50~~~~~~~~~~~~&&1.39~~~~~~~~~~~~\\
log(S/O)                             &--1.503$\pm$0.016~~&&--1.571$\pm$0.068~~\\ \\
Cl$^{++}$/H$^+$($\times$10$^8$)      &1.017$\pm$0.223  &&      ...      \\
ICF(Cl)$^{\rm a}$                          &1.57~~~~~~~~~~~~&&       ...      \\
log(Cl/O)                             &--3.409$\pm$0.076~~&&    ...      \\ \\
Ar$^+2$/H$^+$($\times$10$^7$)        &1.449$\pm$0.032  && 1.743$\pm$0.130\\
Ar$^{+3}$/H$^+$($\times$10$^7$)      &0.608$\pm$0.119  &&      ...      \\
ICF(Ar)$^{\rm a}$                          &1.05~~~~~~~~~~~~&&  1.64~~~~~~~~~~~~\\
log(Ar/O)                             &--2.280$\pm$0.026~~&&--2.158$\pm$0.038\\ \\
Fe$^{++}$/H$^+$($\times$10$^7$)      &1.320$\pm$0.200  &&      ...      \\
ICF(Fe)$^{\rm a}$                          &5.51~~~~~~~~~~~~&&       ...      \\
log(Fe/O)                             &--1.751$\pm$0.016~~&&    ...      \\ 
$[$O/Fe$]$$^{\rm b}$                 &0.331$\pm$0.029  && ...  \\ \\
He$^+$/H$^+$($\lambda$4471)         &0.071$\pm$0.002&& 0.070$\pm$0.009 \\
He$^+$/H$^+$($\lambda$5876)         &0.081$\pm$0.001&& 0.083$\pm$0.004 \\
He$^+$/H$^+$($\lambda$6678)         &0.082$\pm$0.002&& 0.081$\pm$0.008 \\
He$^+$/H$^+$                        &                 &&                  \\
(weighted mean)                     &0.080$\pm$0.001&& 0.080$\pm$0.003 \\
He$^{+2}$/H$^+$($\lambda$4686)      &0.002$\pm$0.000&&       ...       \\
He/H                                &0.082$\pm$0.001&& 0.080$\pm$0.003 \\
$Y$                                 &0.246$\pm$0.003&& 0.243$\pm$0.010 \\ \hline
\end{tabular}

$^{\rm a}$ICF is the ionization correction factor.

$^{\rm b}$$[$O/Fe$]$ $\equiv$ log (O/Fe) -- log (O/Fe)$_\odot$.
\end{table}

\section{Age of the stellar population in the LSB regions \label{age}}

We consider next the spectroscopic and photometric properties of
the LSB regions labeled {\it a}1, {\it a}2 (slit 
position 1 in Fig. \ref{f1}) and {\it a}3, {\it e}3, {\it a}4 (slit position 
2), to constrain the age of the stellar populations 
contributing to the light in those regions.
H$\alpha$ and H$\beta$ emission lines are present in regions {\it a}1, 
{\it a}3 and {\it e}3 while H$\gamma$ and H$\delta$ absorption 
lines are detected in the spectra of all regions except for region {\it e}3.
This
allows us to derive the age of the stellar population using four 
methods, based on: (1) the time evolution of equivalent widths ($EW$)
of hydrogen emission lines, (2) the time evolution of $EW$s of hydrogen
absorption lines, (3) the comparison of the observed 
and theoretical spectral energy distributions, and (4) the broad-band
colours.
The requirement of consistency of the ages determined from the  
reddening-insensitive methods 1 and 2 and from the 
reddening-sensitive methods 3 and 4 allows to simultaneously derive 
the extinction coefficient and constrain the SF history (Guseva et al. 
\cite{Guseva2001,Guseva2003a,Guseva2003b}).

We measured the fluxes and equivalent widths of the 
H$\alpha$ and H$\beta$ emission lines and the H$\gamma$ and H$\delta$ absorption
lines in the spectra of the LSB regions, and list them in 
Tables~\ref{t:emhahb} and \ref{t:abshdhg}.
Because the H$\beta$ emission line is narrower than the absorption line in
these regions and does not fill the absorption component, its flux was 
measured using the continuum level at the bottom of the absorption line.
This level was chosen by visually interpolating from the absorption line
wings to the center of the line.

The extinction coefficient $C$(H$\beta$) in those regions is derived 
from the H$\alpha$/H$\beta$ flux ratio.
We adopt the theoretical recombination H$\alpha$/H$\beta$ flux ratio of
2.8, which is typical for hot low-metallicity H {\sc ii} regions.
No correction for the absorption line equivalent widths has been made.
The extinction coefficients $C$(H$\beta$) are shown in Table \ref{t:emhahb}. 

Hydrogen absorption lines are  seen in the spectra of all 
regions labeled  in Fig.~\ref{f1} except for the  
brightest H {\sc ii} regions {\it e}1 and {\it e}2
and the LSB region {\it e}3.
The blue part of the spectrum of region {\it a}1 with hydrogen absorption 
and emission lines is shown in Fig. \ref{fig:abs}.
Table~\ref{t:abshdhg} lists the equivalent widths with their errors of 
the H$\gamma$ and H$\delta$ absorption lines measured 
in the wavelength  intervals or ``windows'' of Bica \& Alloin (\cite{Bica86}).
The errors include 
the errors in the fitting of line profiles with Gaussians and the noise
dispersion in the
continuum. A careful placement of the continuum level is very important 
for deriving accurate $EW$s.
For this purpose, we choose points in the spectrum free of nebular and stellar lines,
which were then fitted by cubic splines. The uncertainties 
were estimated from several different measurements of the 
equivalent widths of hydrogen absorption lines with independent continuum 
fittings. They are of the same order as the errors in 
Table~\ref{t:abshdhg}. 
 

\begin{table*}[tbh]
\caption{Fluxes, equivalent widths of the H$\alpha$ and H$\beta$ emission 
lines and the extinction coefficients $C$(H$\beta$) in the LSB regions.}
\label{t:emhahb}
\begin{tabular}{lcclrccrrr} \hline \hline
Telescope&Region&Distance$^{\rm a}$&Aperture$^{\rm b}$&& 
$F$(H$\alpha$)$^{\rm c}$  &$F$(H$\beta$)$^{\rm c}$  &$EW$(H$\alpha$)$^{\rm d}$  
&$EW$(H$\beta$)$^{\rm d}$ &$C$(H$\beta$)   \\ \hline

MMT$^{\rm e}$ &{\it a}1& 10.7~~~~~~&1.5$\times$3.4 && 3.53 $\pm$0.14 & 1.25 $\pm$0.12 & 12.30 $\pm$0.75 & 2.72 $\pm$0.24 & 0.0 $\pm$0.03 \\
     &{\it a}2& 29.5~~~~~~&1.5$\times$13.2 && ... & ... &\multicolumn{1}{c}{...}
&\multicolumn{1}{c}{...}&\multicolumn{1}{c}{...} \\ 
  \hline

4m$^{\rm f}$ &{\it a}3& 12.6~~~~~~&2.0$\times$7.6 && 4.88 $\pm$0.39 & 1.88 $\pm$0.29 & 32.81 $\pm$2.32& 7.72 $\pm$1.01 & 0.0 $\pm$0.04 \\
        &{\it e}3& 18.4~~~~~~&2.0$\times$2.8 && 3.69 $\pm$0.17 & 1.19 $\pm$0.14 &109.80 $\pm$1.41& 16.57 $\pm$0.69 & 0.04 $\pm$0.06 \\
        &{\it a}4& 22.6~~~~~~&2.0$\times$2.8 &&     ...       &      ...       &\multicolumn{1}{c}{...}&\multicolumn{1}{c}{...}&\multicolumn{1}{c}{...} \\
 
  \hline
\end{tabular}

$^{\rm a}$distance in arcsec from the brightest H {\sc ii} region {\it e}1. \\
$^{\rm b}$aperture size $x$$\times$$y$, where $x$ is the slit width and $y$ the size 
along the slit in arcsec. \\
       $^{\rm c}$in units 10$^{-16}$\ erg\ s$^{-1}$cm$^{-2}$. \\
$^{\rm d}$in \AA. \\
$^{\rm e}$slit orientation with position angle P.A. = 22$^{\circ}$. \\
$^{\rm f}$slit orientation with position angle P.A. = 48$^{\circ}$. \\
\end{table*}



\begin{table*}[tbh]
\caption{Equivalent widths of the H$\gamma$ and 
H$\delta$ absorption lines in the LSB regions. 
}
\label{t:abshdhg}
\begin{tabular}{lcclrrc} \hline \hline
Telescope &Region&Distance$^{\rm a}$&Aperture$^{\rm b}$&& 
$EW$(H$\delta$)$^{\rm c}$  &$EW$(H$\gamma$)$^{\rm c}$  \\ \hline

MMT$^{\rm d}$ &{\it a}1& 10.7~~~~~~&1.5$\times$3.4 && --5.53 $\pm$0.22  & --4.92 $\pm$0.23 \\ 
     &{\it a}2& 29.5~~~~~~&1.5$\times$13.2  && --8.89 $\pm$0.53 & --6.07 $\pm$0.41 \\ 
  \hline

4m$^{\rm e}$ &{\it a}3& 12.6~~~~~~&2.0$\times$7.6 && --8.28 $\pm$0.50 & --5.71 $\pm$0.60 \\ 
        &{\it a}4& 22.6~~~~~~&2.0$\times$2.8 && --9.06 $\pm$0.63 &     ...       \\ 

  \hline
\end{tabular}

$^{\rm a}$distance in arcsec from the brightest H {\sc ii} region {\it e}1. \\
$^{\rm b}$aperture size $x$$\times$$y$, where $x$ is the slit width and $y$ the size 
along the slit in arcsec. \\
$^{\rm c}$in \AA. \\
 $^{\rm d}$slit orientation with position angle P.A. = 22$^{\circ}$. \\
$^{\rm e}$slit orientation with position angle P.A. = 48$^{\circ}$. \\
\end{table*}


\subsection{Age calibration\label{under}}

The calibration of the age of stellar populations using 
the equivalent widths
of the H$\alpha$ and H$\beta$ nebular emission lines, 
those of the H$\gamma$ and H$\delta$
stellar absorption lines and the spectral energy distributions is 
discussed in 
detail in Guseva et al. (\cite{Guseva2001,Guseva2003a,Guseva2003b}). Here 
we only briefly describe these calibrations.

\subsubsection{Balmer emission lines\label{under_1}}

  \begin{figure*}
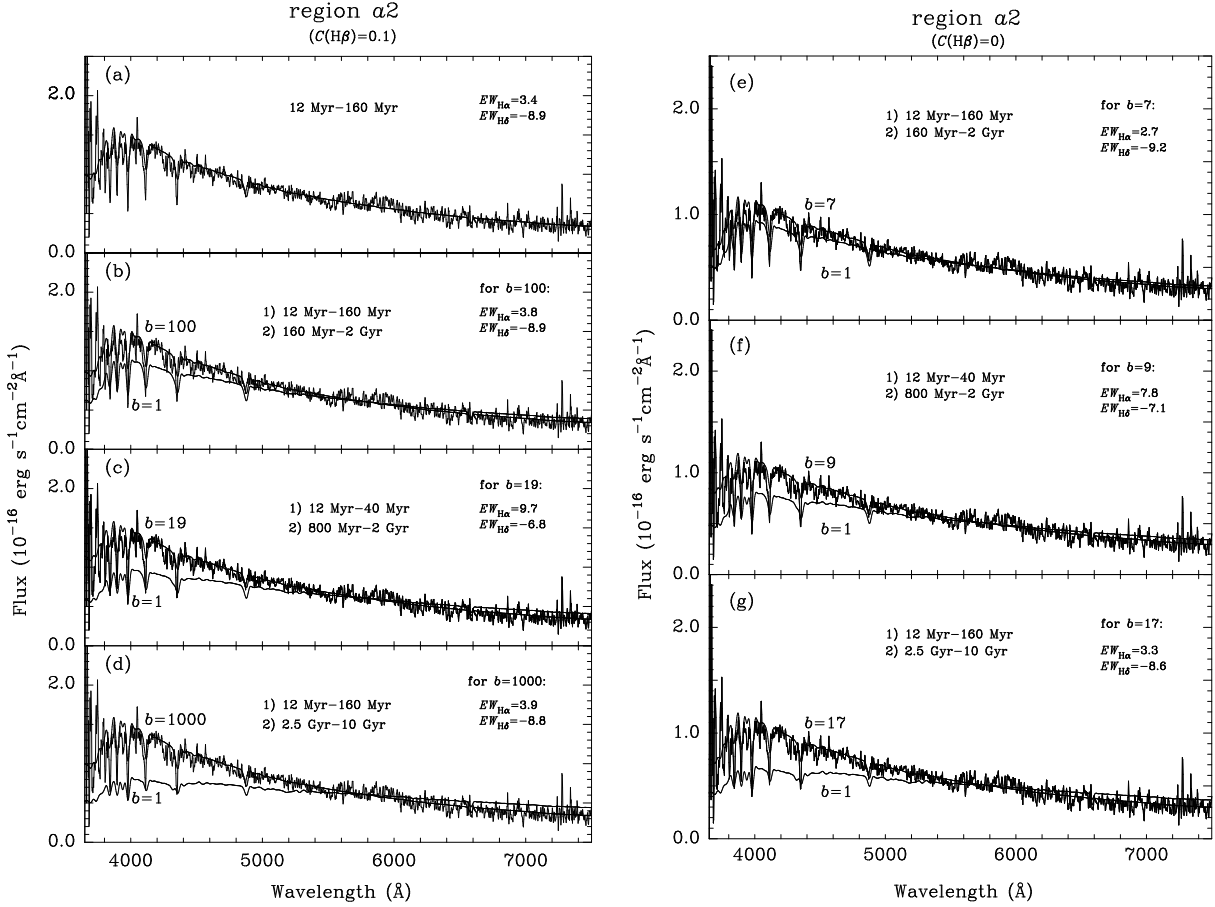

\vspace{1.cm}
    \hspace*{-0.0cm}\psfig{figure=3331.f7a.ps,angle=0,height=12.cm,clip=}
    \hspace*{0.4cm}\psfig{figure=3331.f7b.ps,angle=0,height=11.9cm,clip=}
 \caption[]{Spectra of region {\it a}2 on which 
     synthetic SEDs are superposed. 
     Synthetic SEDs are calculated for stellar populations forming continuously
     during one ({\bf a}) or two ({\bf b} -- 
{\bf g}) time intervals. In the case
     of two intervals the SFR is constant
     within each interval but varies from one interval to another one by a
     factor $b$ = SFR(young)/SFR(old). 
The spectra in left panel are corrected for interstellar
     extinction with $C$(H$\beta$) = 0.1. 
The extinction coefficient $C$(H$\beta$)
     is set to 0 in right panel.
Time intervals, parameters $b$ and 
     predicted $EW$s of the H$\alpha$ emission  and H$\delta$ absorption 
lines are indicated (see explanations in Sect.~\ref{extend}).
    }    
\label{fig:spfit_1}
\end{figure*}

The temporal evolution of the H$\alpha$ and H$\beta$ emission 
line equivalent widths depends on the star formation history.
We consider here the two limiting cases of instantaneous burst
and continuous star formation models. 
The equivalent widths for the instantaneous burst model 
with a heavy element mass fraction $Z_\odot$/20
are calculated
using the galactic evolution code PEGASE.2 (Fioc \& Rocca-Volmerange 
\cite{F97}). The dependence of the H$\alpha$ emission line equivalent
width on time is shown in Fig. 6a of Guseva et al. (\cite{Guseva2003b}) 
by the thick solid line. 
These models are appropriate for regions {\it e}1 and {\it e}2 with strong
emission lines. 
The equivalent widths of Balmer emission lines in region {\it e}1 
($EW$(H$\alpha$) = 998\AA\ and $EW$(H$\beta$) = 166 \AA) 
and region {\it e}2 ($EW$(H$\alpha$) = 872\AA\
and $EW$(H$\beta$) = 134 \AA) correspond to an instantaneous burst 
age of 4 Myr.

However, for the LSB regions, models with continuous
star formation are more appropriate. For these models we adopt 
a constant star formation rate (SFR)
within the time interval from $t_{\rm i}$ when star formation starts to 
$t_{\rm f}$ when it stops. Time is zero now and increases to the past. 
The equivalent widths of hydrogen emission 
lines and SEDs for a set of instantaneous burst models (Fioc \& 
Rocca-Volmerange \cite{F97}) are used to calculate the temporal evolution of
$EW$s for continuous SF with a constant
SFR. The temporal dependence of the equivalent widths of the  
H$\alpha$ emission line is shown in Fig. 6a of Guseva et al. 
(\cite{Guseva2003b}) for different $t_{\rm i}$ and $t_{\rm f}$. 

\subsubsection{Stellar Balmer absorption lines\label{under_2}}

Another way to derive the age of a stellar population is to use
the relation between the H$\delta$ and H$\gamma$ absorption line
equivalent widths and age, derived 
by Gonz\'alez Delgado, Leitherer \& Heckman (\cite{GonLeith99b}). 
Their instantaneous burst models predict a steady increase of the equivalent 
widths with age from 1 Myr to 1 Gyr. However, 
they did not extend the calculations for ages $\ga$ 1 Gyr when the equivalent
widths of the absorption lines decrease with age (Bica \& Alloin 
\cite{Bica86}). Hence, each value of the hydrogen absorption
line equivalent width corresponds to two values of the age, $\la$ 1 Gyr
and $\ga$ 1 Gyr. This ambiguity can be resolved with the use of other age 
constraints discussed in this paper.

 Furthermore, the models by Gonz\'alez Delgado et al. (\cite{GonLeith99b})
probably overestimate the equivalent widths of the absorption lines at ages 
$\sim$ 1 Gyr (Guseva et al. \cite{Guseva2003b}). 
Therefore, in the age range from 1 Myr to 16.5 Gyr instead of the 
calibration by Gonz\'alez Delgado et al. (\cite{GonLeith99b}) 
we use an empirical calibration of the hydrogen absorption line
equivalent widths versus age by Bica \& Alloin (\cite{Bica86}).
This calibration is based on the integrated spectra of 63 open and globular
stellar clusters with 
known ages, metallicities and reddenings
which can be used as templates for 
stellar populations formed in an instantaneous burst.
For consistency we use the same wavelength intervals or ``windows'' 
for H$\gamma$ and H$\delta$ 
flux measurements as Bica \& Alloin (\cite{Bica86}) 
($\lambda$4318--4364\AA\ and $\lambda$4082--4124\AA, respectively).  


\begin{table*}[tbh]
\caption{Predicted equivalent widths of emission and absorption lines,
extinction coefficients and ($V$--$I$) colours from models of continuous star formation.}
\label{models}
\begin{tabular}{lcccrrcrrcccc} \hline \hline
 & &\multicolumn{2}{c}{Age range$^{\rm a}$} && &\multicolumn{4}{c}{modeled $EW$}&&\multicolumn{2}{c}{modeled ($V$--$I$)$^{\rm d}$} \\ \cline{3-4} \cline{7-10} \cline{12-13}
 &Region&young&old&$b$$^{\rm b}$&$M_{\rm y}$/$M_{\rm o}$&H$\delta$$^{\rm c}$&H$\gamma$$^{\rm c}$& 
H$\beta$$^{\rm c}$  &H$\alpha$$^{\rm c}$  
&$C$(H$\beta$)  &intrinsic &reddened \\ 
 & &$t_{\rm f}$ -- $t_{\rm i}$ &$t_{\rm f}$ -- $t_{\rm i}$& & & & & & & & & \\ 
\hline 

MMT$^{\rm e}$ &{\it a}1&\multicolumn{2}{c}{7.1--7.5}&  ... &... &    --6.3& --5.2& 2.2 &   13.8 & 0.27 & 0.211 & 0.452 \\ 
              &        & 7.05--7.5 & 9.4--10 &   24 &0.07&    --6.0& --5.1& 2.5 &   13.2 & 0.00 & 0.456 & 0.456 \\ 
              &        & 7.1--7.5 &  9.4--10 &   66 &0.18&    --6.2& --5.2& 2.2 &   12.9 & 0.12 & 0.335 & 0.442 \\ 
              &        & 7.11--7.5 & 9.4--10 &  400 &1.00&    --6.2& --5.2& 1.9 &   13.6 & 0.22 & 0.242 & 0.438 \\ 
              &        &           &         &      &    &         &      &     &        &      &       &       \\ 
              &{\it a}2&\multicolumn{2}{c}{7.1--8.2}&  ... &... &    --8.9& --7.9& 0.6 &    3.4 & 0.10 & 0.350 & 0.439 \\ 
              &        & 7.1--8.2 & 8.2--9.3 &  100 &8.1 &    --8.9& --7.9& 0.7 &    3.8 & 0.10 & 0.358 & 0.447 \\ 
              &        & 7.1--7.6 & 8.9--9.3 &   19 &0.5 &~\bf --6.8$^{\rm g}$& --5.6& 1.7 &~\,\bf 9.7$^{\rm g}$& 0.10 & 0.342 & 0.431 \\ 
              &        & 7.1--8.2 & 9.4--10  & 1000 &19.7&    --8.8& --7.9& 0.7 &    3.9 & 0.10 & 0.352 & 0.441 \\ 
              &        &          &          &      &    &         &      &     &        &      &       &       \\ 
              &        & 7.1--8.2 & 8.2--9.3 &    7 &0.6 &    --9.2& --7.9& 0.5 &    2.7 & 0.00 & 0.440 & 0.440 \\ 
              &        & 7.1--7.6 & 8.9--9.3 &    9 &0.2 &~\bf --7.1$^{\rm g}$& --5.8& 1.4 &~\bf 7.8$^{\rm g}$& 0.00 & 0.418 & 0.418 \\ 
              &        & 7.1--8.2 & 9.4--10  &   17 &0.3 &    --8.6& --7.7& 0.6 &    3.3 & 0.00 & 0.441 & 0.441 \\ 
  \hline

KPNO 4m$^{\rm f}$  &{\it a}3&\multicolumn{2}{c}{6.7--8.4}& ...  &... &    --8.6& --7.6& 6.6 &   36.8 & 0.10 & 0.355 & 0.444 \\ 
              &        &\multicolumn{2}{c}{6.7--8.8}& ...  &... &~\bf --9.6$^{\rm g}$& --7.8& 4.3 &~\bf 22.6$^{\rm g}$& 0.00 & 0.427 & 0.427 \\ 
              &        & 6.7--8.4 & 8.4--9.3 &    7 &1.0 &    --8.9& --7.6& 5.5 &   29.3 & 0.00 & 0.418 & 0.418 \\ 
              &        & 6.7--8.4 & 9.4--10  &   20 &0.6 &    --8.4& --7.5& 6.3 &   33.0 & 0.00 & 0.412 & 0.412 \\ 
              &        & 6.75--8.4&  9.4--10 & 1000 &32.5&    --8.7& --7.6& 5.4 &   30.5 & 0.12 & 0.361 & 0.468 \\ 
              &        &          &          &      &    &         &      &     &        &      &       &       \\ 
              &{\it a}4&\multicolumn{2}{c}{7.1--8.2}& ...  &... &    --8.9& --7.9& 0.6 &   3.4  & 0.00 & 0.350 & 0.350 \\ 
              &        & 7.1--8.2 & 8.2--9.3 &  100 &8.1 &    --8.9& --7.9& 0.7 &   3.8  & 0.00 & 0.358 & 0.358 \\ 
              &        & 7.1--7.6 & 8.9--9.3 &   19 &0.5 &~\bf --6.8$^{\rm g}$& --5.6& 1.7 &~\,\bf 9.7$^{\rm g}$& 0.00 & 0.342 & 0.342 \\ 
              &        & 7.1--8.2 & 9.4--10  & 1000 &19.7&    --8.8& --7.9& 0.7 &   3.9  & 0.00 & 0.352 & 0.352 \\ 

  \hline
\end{tabular}

$^{\rm a}$duration of SF in log $t$ ($t$ in yr). \\ 
$^{\rm b}$$b$ $\equiv$ SFR(young)/SFR(old). \\
$^{\rm c}$$EW$ in \AA. \\
$^{\rm d}$intrinsic colour is that for the stellar population in the absence of extinction. $(V-I)_{\rm reddened}$ =
$(V-I)_{\rm intrinsic}$ + 0.891 $C$(H$\beta$). \\
$^{\rm e}$slit orientation with position angle P.A. = 22$^{\circ}$. \\
$^{\rm f}$slit orientation with position angle P.A. = 48$^{\circ}$. \\
$^{\rm g}$$EW$s do not fit observations. \\
\end{table*}

\begin{figure*}
\begin{picture}(16,11)
\put(2,0){{\psfig{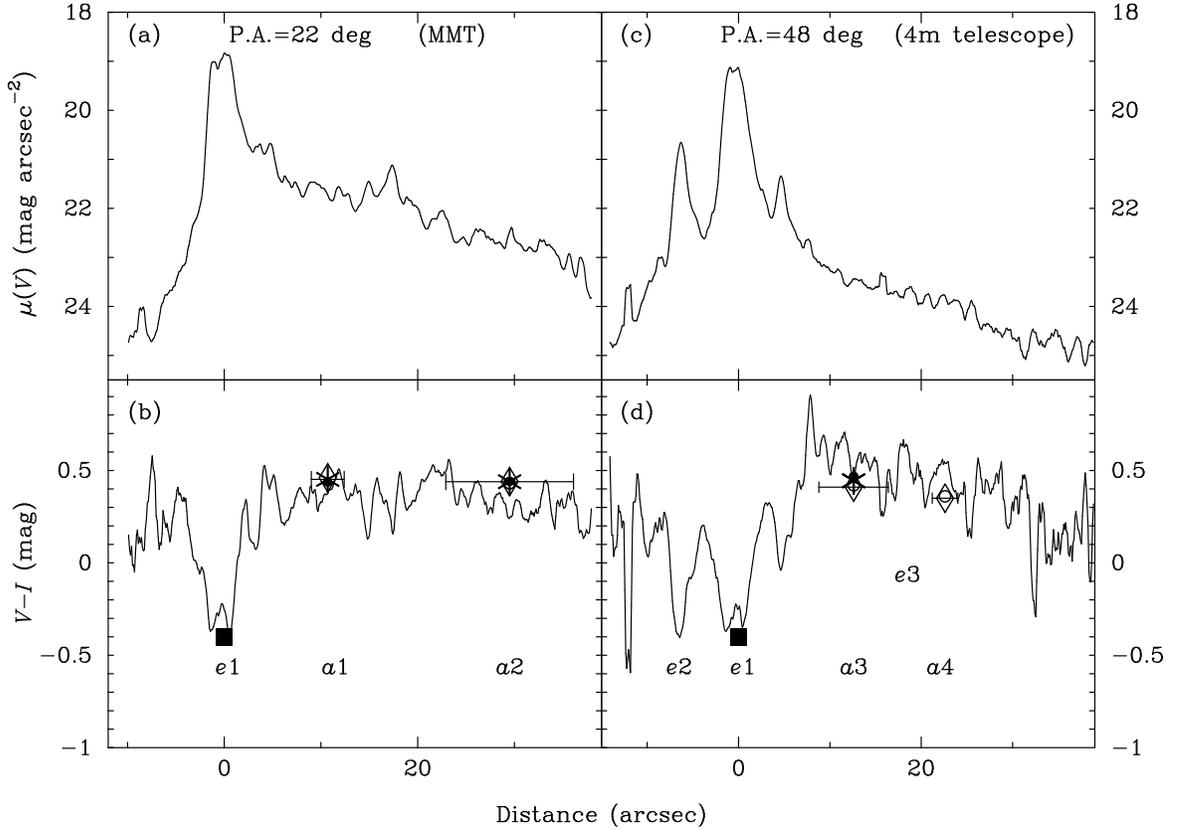}}}
\end{picture}
\caption[]{({\bf a}) $V$ surface brightness distribution
along the slit oriented at position angle P.A. = 22$^\circ$. 
The origin is set at the brightest H {\sc ii} region {\it e}1.
({\bf b}) ($V-I$) colour distribution along the slit with position angle
P.A. = 22$^\circ$. 
({\bf c}) $V$ surface brightness distribution along the
slit oriented at position angle P.A. = 48$^\circ$. 
The origin is set at the brightest H {\sc ii} region {\it e}1.
({\bf d}) ($V-I$) colour distribution along the slit with position angle
P.A. = 48$^\circ$. The locations of different regions are 
labeled in ({\bf b}) and ({\bf d}).
Filled squares show the total modeled
colours (stellar plus gaseous emission) for region {\it e}1.
Other symbols show the colours predicted for the LSB regions by the  
models considered (Table~\ref{models}). 
  } 
\label{f6}
\end{figure*}

  \begin{figure}[hbtp]
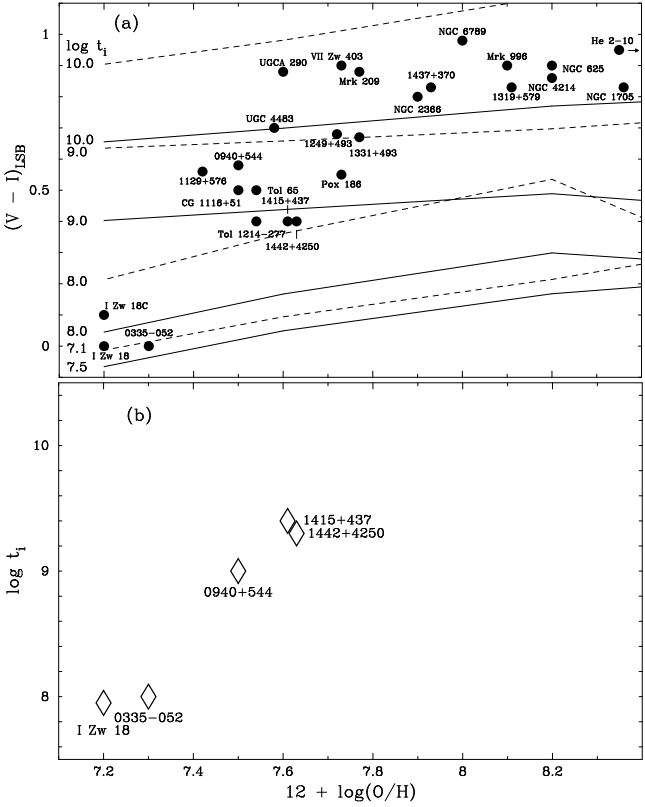

    \psfig{figure=3331.f9a.ps,angle=270,width=8.5cm}
    \hspace*{0.03cm}\psfig{figure=3331.f9b.ps,angle=270,width=8.48cm}
    \caption{Dependence of the ($V-I$) colours of the LSB component  ({\bf a})
     and logarithm of the derived ages ({\bf b})
     on the oxygen abundance for the dwarf irregular and BCD galaxies from our sample.
    Dashed lines in ({\bf a}) show theoretical dependences of 
    the ($V$--$I$) colour on the oxygen abundance for an instantaneous burst calculated 
    using the galactic evolution code PEGASE.2 (Fioc $\&$ 
    Rocca-Volmerange 1997) in the range of ages (in logarithmic scale)
    log $t_{\rm i}$ between 7.1 and 10.0 ($t_{\rm i}$ in yr). 
    Solid lines show models in which stars are forming 
    continuously at a constant SFR between $t_{\rm f}$=0 and different $t_{\rm i}$, where
    log $t_{\rm i}$ varies between 7.5 and 10.0. 
    The observed colours are corrected only for reddening in our Galaxy.
    The ages in ({\bf b}) are from Izotov et al. (\cite{ICFTGPFG01b}) (I Zw 18),
    Papaderos et al. (\cite{Papa98}) (SBS 0335--052), Guseva et al. (\cite{Guseva2001})
    (SBS 0940+544), Guseva et al. (\cite{Guseva2003b}) (HS 1442+4250) and this paper
    (SBS 1415+437).
         }
    \label{fig:comp_OH}
\end{figure}

The behaviour of the empirical H$\delta$ absorption line equivalent width 
with the age for an instantaneous burst (Bica \& Alloin 
\cite{Bica86}) is shown in Fig.~6b of Guseva et al. (\cite{Guseva2003b})
by the thick solid line.

The temporal evolution of the H$\gamma$ and H$\delta$ 
absorption line equivalent widths in the case of continuous SF
is calculated similarly to that of the H$\alpha$ and H$\beta$ 
emission line equivalent widths described in the section \ref{under_1}. 
More specifically, we use the empirical equivalent widths of hydrogen 
absorption lines (Bica \& Alloin \cite{Bica86})
and SEDs for instantaneous bursts  
(Fioc \& Rocca-Volmerange \cite{F97})
to calculate the temporal evolution of $EW$s in the case of continuous 
SF with constant SFR. The results are shown in 
Fig. 6b of Guseva et al. (\cite{Guseva2003b}) for
SF with different $t_{\rm i}$ and $t_{\rm f}$.

\subsubsection{Spectral energy distribution \label{SED}}

The shape of the spectrum reflects the properties of the
stellar population. However, it is also dependent on the reddening.
A precise determination of the extinction can be done only for the
two brightest H {\sc ii} regions {\it e}1 and {\it e}2 which possess 
many strong hydrogen emission lines (Table~\ref{t:Intens}). 
We derived an extinction coefficient $C$(H$\beta$)=0 in these regions. 
In the LSB regions {\it a}1, {\it a}3 and
{\it e}3,  only H$\alpha$ and H$\beta$ emission lines are present. 
The extinction coefficients obtained from the H$\alpha$/H$\beta$ flux ratio in 
these regions are small (Table \ref{t:emhahb}). However, they are more uncertain as 
compared to the ones in regions {\it e}1 and {\it e}2 because of the weakness
of the emission lines and significant contribution of the stellar
absorption lines. 
 H$\alpha$ and H$\beta$ emission lines are not detected in the other LSB regions. 
Therefore, the  observed SED cannot directly give information on the age,
but should be used together with the methods discussed in Sect. \ref{under_1} 
and \ref{under_2} for simultaneous determination of the age and 
interstellar extinction.

We used the galactic evolution code PEGASE.2 (Fioc \& 
Rocca-Volmerange \cite{F97}) to produce a grid of theoretical SEDs
for an instantaneous burst of star formation with ages ranging between  
0 and 10 Gyr, and a heavy element mass fraction of $Z$ =  $Z_\odot$/20.
The SEDs for continuous SF in the time interval between $t_{\rm i}$  ago and
$t_{\rm f}$ ago are derived by integration of instantaneous burst SEDs.


\begin{table*}[tbh]
\caption{($V-I$) colours of the extended LSB components in some irregular and BCD galaxies.
}
\label{t:com}
\begin{tabular}{lllcccllrr} \hline \hline
No.&Galaxy && ($V-I$)$_{LSB}$$^{\rm a}$  &Telescope$^{\rm b}$&&12+log(O/H)&Telescope$^{\rm c}$&Ref.$^{\rm d}$ &Ref.$^{\rm e}$
\\ \hline

1 & NGC 6789     && 0.98      & {\sl HST}  && 7.7 or 8.5     & --   &${\rm 19}$ &${\rm 2}$  \\
2 & He 2-10      && 0.95      & {\sl HST}  && 7.9$^{\rm ***}$ & 4m &${\rm 19}$ &${\rm 11}$  \\
3 & NGC 625      && 0.90      & {\sl HST}  && $\sim$8.2$^{\rm *}$& 1.6m &${\rm 19}$ &${\rm 16}$  \\ 
4 & Mrk 996      && 0.90 & {\sl HST}  && 8.1            & {\sl HST}  &${\rm 15}$&${\rm 15}$ \\
5 & VII Zw 403   && 0.90 & {\sl HST}  && 7.73           & 4m   &${\rm 13}$&${\rm 9}$  \\  
6 & Mrk 209      && 0.88      & 2.1m && 7.77           & 4m   &${\rm 19}$ &${\rm 8}$  \\
7 & UGCA 290     && 0.88      & {\sl HST}  && $\sim$7.6      & {\sl HST}  &${\rm 19}$ &${\rm 1}$  \\
8 & NGC 4214     && 0.86      & {\sl HST}  && 8.2            & 2.1m &${\rm 19}$ &${\rm 10}$  \\
9 & SBS 1437+370 && 0.83      & 2.1m && 7.93           & 4m   &${\rm 19}$ &${\rm 8}$  \\
10& SBS 1319+579 && 0.83      & 2.1m && 8.11           & 4m   &${\rm 19}$ &${\rm 8}$  \\
11& NGC 1705     && 0.83      & {\sl HST}  && 8.36           & 1m   &${\rm 19}$ &${\rm 14}$ \\
12& NGC 2366     && 0.80 & {\sl HST}  && 7.9            & 4m   &${\rm 19}$ &${\rm 8}$ \\
13& UGC 4483     && 0.70 & {\sl HST}  && 7.58           & 4m   &${\rm 7}$ &${\rm 8}$  \\
14& SBS 1249+493 && 0.68      & 2.1m && 7.72           & 4m   &${\rm 19}$ &${\rm 8}$  \\ 
15& SBS 1331+493 && 0.67      & 2.1m && 7.77           & 4m   &${\rm 19}$ &${\rm 8}$  \\ 
16& SBS 0940+544 && 0.58      & 2.1m && 7.50           & Keck &${\rm 3}$ &${\rm 3}$ \\
17& SBS 1129+576 && 0.56& 2.1m   && 7.42           & 4m &${\rm 17}$ &${\rm 17}$ \\
18& Tol 65       && 0.50 & {\sl HST}  && 7.54           & Keck &${\rm 19}$ &${\rm 4}$ \\
19& CG 1116+51   && 0.50       & 2.1m && 7.5            & MMT  &${\rm 19}$ &${\rm 19}$ \\
20& HS 1442+4250&& 0.40 & 2.1m && 7.63           & 4m   &${\rm 18}$ &${\rm 18}$ \\
21& SBS 1415+437 && 0.40 & 2.1m && 7.61           & 4m   &${\rm 19}$ &${\rm 19}$ \\
22& Tol 1214-277 && 0.40 & {\sl HST}  && 7.54           & Keck &${\rm 19}$ &${\rm 4}$ \\
23& Pox 186      && 0.40 & {\sl HST}  && 7.73           & MMT  &${\rm 19}$ &${\rm 19}$ \\ 
24& I Zw 18C     && 0.10      & {\sl HST}  && 7.2$^{\rm **}$& Keck &${\rm 5}$ &${\rm 5}$ \\
25& SBS 0335-052 && 0.0       & {\sl HST}  && 7.30           & Keck &${\rm 12}$&${\rm 6}$ \\
26& I Zw 18      && 0.0       & 2.1m, {\sl HST}&& 7.2       & Keck &${\rm 13}$&${\rm 6}$ \\     

  \hline
\end{tabular}

$^{\rm a}$colour corrected for reddening in our Galaxy. $E$($V-I$)$_{\rm Galaxy}$ are taken from the NED. \\ 
$^{\rm b}$telescope used to obtain photometry. \\
$^{\rm c}$telescope used to obtain spectroscopic data.  \\
$^{\rm d}$reference for the  photometry.  \\
$^{\rm e}$reference for the chemical abundance. \\ 
$^{\rm *}$our rough estimate. \\
$^{\rm **}$oxygen abundance of the main component I Zw 18 NW + SE. \\
$^{\rm ***}$oxygen abundance is calculated using the observed data by 
Kobulnicky et al. (1999) and 
calibration by Pilyugin (2000).  \\
References: (1) Crone et al. 2002; 
(2) Drozdovsky et al. 2001;
(3) Guseva et al. 2001; 
(4) Izotov et al. 2001a; 
(5) Izotov et al. 2001b; 
(6) Izotov et al. 1999;
(7) Izotov \& Thuan 2002; 
(8) Izotov \& Thuan 1999; 
(9) Izotov et al. 1997a; 
(10) Kobulnicky \& Skillman 1996;  
(11) Kobulnicky et al. 1999;
(12) Papaderos et al. 1998; 
(13) Papaderos et al. 2002; 
(14) Storchi-Bergmann et al. 1994; 
(15) Thuan et al. 1996; 
(16) Vaceli et al. 1997; 
(17) Guseva et al. 2003a; 
(18) Guseva et al. 2003b; 
(19) this paper.  \\
\end{table*}

\subsection{Ages of the stellar populations in the LSB regions\label{extend}}

In this section we derive self-consistently the ages of the stellar populations 
in the LSB regions using:
1) the equivalent widths of emission lines, 2) the equivalent widths of 
absorption lines, 3) the SEDs and 4) the colours. 
For this we adopt a continuous
SF scenario with constant or variable SFR. In the latter case we 
consider a simplified scenario  with two time intervals of SF, which we
call young and old, with
different SFRs. To quantify the difference in SFRs we introduce the parameter 
$b$ = SFR(young)/SFR(old). Because of the noisy spectrum of region 
{\it e}3 and the significant contamination by nebular emission we were not able 
to measure equivalent widths of absorption lines
in that region. Therefore, we exclude it from the analysis. For the
remaining LSB regions, the model predictions providing the best fits to the observed 
SEDs are shown in Table \ref{models}. For each region we show the
age range of one or two SF episodes, the parameter $b$, the mass ratio 
$M_{\rm y}$/$M_{\rm o}$ of the young-to-old populations,
the model equivalent widths of hydrogen lines, 
the extinction coefficients $C$(H$\beta$) derived from the best
match between observed and calculated SEDs,
and the intrinsic and reddened ($V-I$) colours (see Sect.~\ref{coldist}). 
We use the 
relation $E$$(V-I)$ = 0.891 $C$(H$\beta$) (e.g.,
Aller \cite{Aller84}) to correct for reddening. 
Our young and old stellar populations in Table~\ref{models} include stars 
with ages not older than 250 Myr (log $t$ = 8.4) and not younger than 160 Myr 
(log $t$ = 8.2), respectively. These definitions differ from those conventionally used.
In fact, the old stellar population in our case includes not only several 
Gyr old stars but also intermediate-age stars with age $\la$ 1 Gyr.  
Negative $EW$s denote absorption lines,
positive $EW$s refer to emission lines. Models with highlighted $EW$s are
those in which the equivalent
widths are not reproduced well despite a good fit of the observed SEDs
(compare model $EW$s in Table~\ref{models} with observed $EW$s in
Tables~\ref{t:emhahb} and ~\ref{t:abshdhg}). 

We show in Fig.~\ref{fig:spfit_1}  
the observed 
and predicted SEDs in the outer LSB region {\it a}2 (MMT 
observations) with no emission features. 
The  H$\gamma$ and H$\delta$ absorption 
lines in this region as well as in the other outer  
region {\it a}4 (4m telescope observations)
 are not contaminated by nebular emission from young stellar populations.
  Hence, 
the absorption line equivalent widths in regions 
{\it a}2 and {\it a}4 are more accurate than in other LSB regions 
(Table~\ref{t:abshdhg}).

We consider several SF histories  and vary 
extinction in the LSB regions to put constraints on the age of their stellar 
populations. 
Figure \ref{fig:spfit_1} shows the spectra of region 
{\it a}2 on which are superposed the synthetic SEDs with different SF 
histories which best fit it. 

We have assumed two extinction coefficients: 
$C$(H$\beta$) = 0.1, derived 
by Thuan et al. (1999), and corresponding to $A(V)$ $\sim$ 0.2, 
and $C$(H$\beta$) = 0.

The model SEDs, adopting $C$(H$\beta$) = 0.1 are
shown in Fig.~\ref{fig:spfit_1}, left panel.
The observed properties of region {\it a}2 can be reproduced by a 
single  young stellar
population with age between 12 Myr and 160 Myr, if a small amount of extinction
is assumed (Fig.~\ref{fig:spfit_1}a). 
There is no need to invoke an older stellar population which,
if present, has to be much smaller in mass than the young stellar population.
We note that there is no ongoing SF in this region.
The most recent SF in region {\it a}2
stopped at time  $\ga$ 12 Myr ago, otherwise 
its spectrum would have shown
a detectable H$\alpha$ emission line (Fig. 6a in 
Guseva et al. \cite{Guseva2003b}). 

Next we consider SF scenarios with a varying SFR which include
older stellar populations with an age of up to 2 Gyr 
(Fig.~\ref{fig:spfit_1}b --\ref{fig:spfit_1}c)  and 10 Gyr 
(Fig.~\ref{fig:spfit_1}d). But even in these cases, the young 
population completely dominates the
light and mass of  region {\it a}2 (Table~\ref{models}). 

An  upper limit to the age of region {\it a}2 can be obtained by assuming 
no extinction, i.e. $C$(H$\beta$) = 0 (Fig.~\ref{fig:spfit_1}, 
right panel). 
Then the observed SED cannot be reproduced by a
synthetic SED with stars forming between 12 Myr and 160 Myr ago, as the 
latter is  too blue. However, by varying the parameter $b$, the observed 
SED can be reproduced by SEDs of stellar populations with other SF scenarios. 
In Figs.~\ref{fig:spfit_1}b -- \ref{fig:spfit_1}d, 
\ref{fig:spfit_1}e -- \ref{fig:spfit_1}g, we show the best 
fits labeled by the adopted value of $b$.
We exclude the models shown in 
Fig.~\ref{fig:spfit_1}c and ~\ref{fig:spfit_1}f because 
they do not fit 
the observed equivalent widths of H$\alpha$ and H$\delta$. The remaining two
models with $C$(H$\beta$) = 0 satisfy all observational constraints. 
In these models, the SFRs for the young population are respectively
7 times (Fig. \ref{fig:spfit_1}e) and 17 times 
(Fig. \ref{fig:spfit_1}g) larger than SFRs for the old population. 
These ratios are, however,  
 significantly smaller than the corresponding parameters $b$ in the case with
$C$(H$\beta$) = 0.1. The relative mass fraction of the young stellar
population is therefore smaller in the extinction-free case, 
being $M_{\rm y}$/$M_{\rm o}$ = 0.6 (Fig.~\ref{fig:spfit_1}e) 
and 0.3 (Fig.~\ref{fig:spfit_1}g),
respectively. Therefore, the presence of a 2 Gyr old or even 10 Gyr old 
population is not excluded in region {\it a}2, which would however  not
dominate the optical emission. If the extinction is non-zero, which is
likely the case, then there is no need to invoke a significant old population
to explain the observed properties of region {\it a}2. It is likely that
for other regions, {\it a}1 and {\it a}3 (Table \ref{models}), the extinction 
is even larger. 

Similar consideration for region {\it a}4 shows that the mass fraction 
of the old stellar population  is small even in the absence of 
extinction. In any case, the old stellar population, if present 
does not contribute significantly to the luminosity. All 
the observed properties of this region can 
be reproduced with a young stellar population formed between 12 
Myr and 160 Myr ago. However, the presence of an older stellar 
population cannot be excluded in that case, but one 
needs to increase the parameter $b$ to match the
observations (Table~\ref{models}).

If some extinction is present in region {\it a}4, then our age 
upper limits will decrease.
There is some hint that the extinction may increase with 
decreasing distance to the brightest H {\sc ii} region {\it e}1. Indeed, 
by considering models which include also  a 10 Gyr old  population and 
$C$(H$\beta$) = 0, 
we find  from Table \ref{models} that the closer a region is
to the brightest part of the galaxy
(region {\it a}1 compared to region {\it a}2, region {\it a}3 compared to 
region {\it a}4), the smaller is the relative mass $M_{\rm y}$/$M_{\rm o}$
of the young stellar population. This is in contrast to the trend 
found in BCDs
where the relative mass of the young stellar population decreases
outwards. Hence, we conclude, that the extinction in SBS 1415+437 is
larger in the inner brighter regions.

 In summary, our spectroscopic analysis of the LSB regions 
favors a relatively young luminosity-weighted age of the stars 
populating those regions.
A model with a constant SFR continuing 
 from 10 Gyr ago until now is definitely excluded. 
An old  population $\ga$ 2 Gyr could be present only in models with
very specific SF scenarios, with a very low SF activity during
the first 2--10 Gyr period, a very high star formation rate during the last
$\sim$(10--200) Myr, and a quiescent period in between. If, however, low SF 
activity has occurred in the period between $\sim$200 Myr and 2 Gyr ago,
then there is no need to invoke a stellar population with age 
$\ga$ 2 Gyr, and all the spectroscopic properties of the LSB regions 
can be explained by only young and intermediate-age
stellar populations.

\subsubsection{Age from the colour distributions 
\label{coldist}}

From the {\sl HST} images, we derived $V$ and $I$ surface brightness and colour 
distributions  for the regions covered by the spectroscopic observations 
with the MMT (Fig.~\ref{f6}a and ~\ref{f6}b) and the KPNO 4m telescope 
(Fig. \ref{f6}c and ~\ref{f6}d). Note the trend for the  ($V-I$) colour 
of the LSB component to decrease  with
increasing distance from region {\it e}1, especially for the KPNO 4m data (Fig. \ref{f6}d). 
This is  again suggestive of larger extinction in the brighter regions.
We compare the observed colours with predictions from our
population synthesis modeling. The results of this comparison are shown in
Figs.~\ref{f6}b and \ref{f6}d. The predicted colours,
obtained from convolving the theoretical SEDs
with the appropriate filter bandpasses, are shown by different symbols.
The transmission curves for the Johnson $V$ and Cousins $I$ bands are taken
from Bessell (\cite{B90}). The zero points are from Bessell, Castelli \& Plez
(\cite{B98}). 

Since the contribution of ionized gas emission to the total brightness of 
region {\it e}1 is significant, 
the theoretical SED for this region has been constructed  
using a 4 Myr old stellar population SED for a heavy element mass fraction
$Z$ = $Z_\odot$/20 to which the gaseous continuum SED and the
observed emission lines were added  
(see Guseva et al. \cite{Guseva2001}).
For comparison with the observed colour at P.A. = 22$^{\circ}$, we reddened 
the predicted colour of region 
{\it e}1 by an amount corresponding to $C$(H$\beta$) = 0.11 
(Table 3 in Thuan et al. \cite{ti99}),
and at P.A. = 48$^{\circ}$ we 
adopted $C$(H$\beta$) = 0 (Table~\ref{t:Intens}).
 The observed colour of region {\it e}1 is very blue, ($V$ -- $I$) 
$\sim$ --0.4 (Fig.~\ref{f6}b and ~\ref{f6}d), and cannot be 
reproduced by a 4 Myr old stellar population alone (($V$ -- $I$) $\sim$ 0.1). 
Strong gaseous continuum and line emission need to be added
(Table~\ref{t:Intens}).  

On the other hand, in all the LSB regions the contribution of the gaseous 
emission to the total flux is negligible. 
In Figs.~\ref{f6}b and  \ref{f6}d we show by different symbols the
colours of the LSB regions with various star formation 
history (Table~\ref{models}).
The total colour (stellar plus gaseous emission) of region {\it e}1 
is shown by a filled squares.
The agreement between the ($V-I$)  colours
and those derived from the spectral energy
distributions is very good.
However, some uncertainties in the  colours may be introduced by
the uncertainties in the reddening curves and the ratios of total to selective
extinction $R$=$A(V)$/$E(B-V)$. We use $R$ = 3.2 by Aller (1984), based
on the reddening curve by Seaton (1979).
Schlegel, Finkbeiner \& Davis (1998) give a slightly different $R$ = 3.315.
 Nevertheless, because the extinction in the studied regions is small, the errors 
introduced by the use of different reddening curves and $R$
are less than 2\%\ in the $V$ band and negligible in the $I$ band.
The ($B-V$) colour of $\sim$ 0.2 mag, derived from the 
SEDs of the LSB regions, is also consistent with the observed value, 
derived from the $B$ and $V$ SBPs, which are shown in Fig.~\ref{sbp}a.

 \subsection{Comparison of the LSB component photometric properties for 
galaxies with different metallicities \label{comp}}

 Our comprehensive studies of selected galaxies 
with  oxygen abundances 12 + log(O/H) $\la$ 7.6 and
 blue LSB components 
(SBS 0335--052, Izotov et al. \cite{ILCFGK97b}, Papaderos et al. \cite{Papa98}; 
I Zw 18, Izotov et al. \cite{ICFTGPFG01b}, Papaderos et al. \cite{papaderos02}; 
SBS 0940+544, Guseva et al. \cite{Guseva2001};
Tol 1214--277, Fricke et al.  \cite{Fricke01};
Tol 65, Papaderos et al. \cite{Papa99};
SBS 1129+576,  Guseva et al. \cite{Guseva2003a};
HS 1442+4250,  Guseva et al. \cite{Guseva2003b}; 
\sbs, this paper) have led us to  the conclusion that these galaxies might 
be young. This is in contrast 
to the large age estimates for some well-studied higher-metallicity 
irregular and BCD galaxies, such as VII Zw 403 and
UGCA 290 (Schulte-Ladbeck et al. 1998; Crone et al. \cite{Crone02}).
 To investigate this apparent inconsistency  we therefore 
compare  the ($V$--$I$) colours of the LSB components (($V$--$I$)$_{LSB}$)  
of the galaxies from our 
sample with those in galaxies where a large age was derived from  
colour-magnitude diagrams (CMD).
  
  In Table~\ref{t:com}, we show the oxygen abundances and  ($V$--$I$) colours of the 
LSB components in several dwarf irregular and BCD galaxies.
The telescopes used for 
photometric and spectroscopic observations are also given in the Table.
Ground-based photometric data in $V$ and $I$ were obtained with the 2.1m
KPNO telescope by Y. Izotov and R. Green.  The {\sl HST} photometric data
were retrieved from the archive of the Space Telescope Science Institute 
(STScI)\footnote{STScI is operated by the AURA, Inc.,  
under NASA contract NAS5-26555.}.
We included in the sample only galaxies with low internal and/or foreground
interstellar extinction to avoid uncertainties introduced by the 
correction of the ($V-I$) colour for reddening.
Thus, we do not include the nearby star-forming galaxy NCG 1569 
studied with {\sl HST}, for example. 
The chemical abundances for the majority of galaxies are
obtained from spectroscopic observations of 
their H {\sc ii} regions. An exception 
is I Zw 18C, the  faint  component of the BCD I Zw 18, where 
no emission lines of  heavy elements  
were detected.
Therefore, for I Zw 18C we adopt the oxygen abundance 
derived for the bright main body of I Zw 18.
($V-I$) colours of the galaxies in Table~\ref{t:com} are corrected for 
reddening in our Galaxy, with $E$($V-I$)$_{\rm Galaxy}$ 
taken from the NASA/IPAC Extragalactic Database (NED).

  The  dependence of the ($V$--$I$)$_{LSB}$ colours 
 on oxygen abundance for 26 selected galaxies 
is shown in Fig.~\ref{fig:comp_OH}a.
Thin dashed lines represent theoretical dependences 
 for an instantaneous burst 
in the age range from log $t_{\rm i}$ = 7.1 to log $t_{\rm i}$ = 10.0 
($t_{\rm i}$ in yr). 
The models for continuous star formation are shown
by thick solid lines.
These models are calculated for a constant SFR which started at time $t_{\rm i}$, with log $t_{\rm i}$ between 7.5 and 10.0 and continuing until now 
($t_{\rm f}$ = 0). The blueing of the ($V-I$) colour with decreasing 
oxygen abundance is in agreement with model predictions. However, this blueing 
trend for galaxies with 12+log(O/H) $\la$ 7.6 is too
steep to be explained only by metallicity effects. 
The blue colours of the LSB components of these low-metallicity
BCDs 
are also not due to ionized gas emission, as the latter
dominates the outer parts of only two BCDs: I Zw 18
(Izotov et al. 2001b; Papaderos et al. 2002) and SBS 0335--052 (Izotov et al. 1997b;
Papaderos et al. 1998). In other galaxies,
including I Zw 18C, the LSB component emission has mainly a stellar origin.
The steep trend  cannot be explained by
 reddening effects because the interstellar extinction derived 
from the spectroscopic observations is small for the galaxies 
shown in Fig.~\ref{fig:comp_OH}a.
Therefore, it is likely that the blueing is mainly due to a change in
the age of the stellar populations, and that
low-metallicity galaxies are younger than high-metallicity ones.
All objects with 12+log(O/H) $\la$ 7.6 and ($V$--$I$)$_{LSB}$ $\la$ 0.6 mag 
are in the range of colours predicted
for ages $\la$1--2 Gyr by instantaneous and continuous 
models of star formation. On the other hand, galaxies with  
($V$--$I$)$_{LSB}$ $\ga$ 0.7 mag are likely to be older,
with ages $\sim$10 Gyr. Yi (2003) 
has analysed the uncertainties in the synthetic 
integrated ($V-I$) colours caused by uncertainties in the stellar 
evolutionary models, population synthesis techniques and  stellar atmosphere 
models.
He has shown that, despite all the uncertainties, 
a ($V$--$I$) $\la$ 0.65--0.75 mag can be 
attributed  to an intermediate-age stellar population with an age not larger 
than 2 Gyr, in the case of an instantaneous burst of star formation,
in agreement with the age derived here from the PEGASE.2 models 
(dashed lines in Fig.~\ref{fig:comp_OH}a). 
The smaller luminosity-weighted age of galaxies with blue
LSB components in Fig.\ref{fig:comp_OH}a is supported by the detailed
analysis of the spectroscopic and photometric properties of some of these
galaxies, as demonstrated in this series of papers.

Izotov \& Thuan (\cite{IT99}) have suggested that the oxygen abundance may
be a good age indicator. That this appears to be the case is shown in 
Fig.~\ref{fig:comp_OH}b where a clear trend of increasing age of the LSB 
stellar population with increasing  oxygen 
abundance is seen for 5 star-forming galaxies with available data.
The ages for 4 galaxies (I Zw 18, Izotov et al. \cite{ICFTGPFG01b};
SBS 0940+544, Guseva et al. \cite{Guseva2001}; HS 1442+4250, Guseva et al.
\cite{Guseva2003b}; SBS 1415+437, this paper) are
determined by the four methods described before. The age for SBS 0335--052
(Papaderos et al. \cite{Papa98}) is derived using only colours and SEDs.
This is because its LSB component is embedded into the H {\sc ii} region produced 
by the young central clusters and no absorption lines are present in the spectra.

 \section{Conclusions \label{conc}}

The results of 
a detailed photometric and spectroscopic study
of the metal-deficient blue compact dwarf galaxy SBS 1415+437
are presented. H$\alpha$ images and spectra in the optical range have been obtained 
with the Kitt Peak 2.1m and 4m telescopes, respectively. 
A $B$ image has been obtained with the 2.2m Calar Alto telescope.
These data are
supplemented by {\sl HST}/WFPC2 $V$ and $I$ images and MMT spectra from
Thuan et al. (\cite{ti99}).
The main conclusions of this study can be summarized as follows:

\begin{enumerate}

\item SBS 1415+437 is a nearby ($D$ = 11.4 Mpc)  low-metallicity  cometary 
BCD with two bright H {\sc ii} regions in the SW part of the elongated 
low-surface-brightness (LSB) stellar component.
The scale lengths $\alpha$ of the LSB component, obtained from 
surface brightness profiles (SBPs)  
 in $V$ and $I$  are both $\sim$ 0.3 kpc, in excellent agreement with those in 
Thuan et al. (\cite{ti99}).
The observed ($V-I$) colour of the brightest
H {\sc ii} region  is very blue, $\sim$ --0.4 mag, due to the combined effects of a
young stellar population and ionized gas emission. The colours of the 
low-surface-brightness  component 
are much redder ($\sim$ 0.4--0.5 mag) and roughly constant in the
outer parts of the galaxy.
A deeper $B$ band SBP of \sbs\ reveals at intermediate and large radii two
exponential intensity regimes with a scale length of 0.3 and 0.27 kpc, respectively.
The inner one ($4\arcsec \la R^* \la 13\arcsec$) can be identified 
with the exponential component derived from {\it HST}~$V$ and $I$ data.
The outer one, which dominates for $R^*$ $\ga$ 16\arcsec, is
$\ga$ 0.3 mag fainter than the inner component.

\item In the two brightest H {\sc ii} regions we derive
 oxygen abundances of 12 + log(O/H) = 7.61 $\pm$ 0.01 and 
7.62 $\pm$ 0.03 ($Z$ $\sim$ $Z_\odot$/20). 
These values agree well with previous determinations (Thuan et al. \cite{ti99}).
Other heavy element-to-oxygen abundance 
ratios for these H {\sc ii} 
regions are also in good agreement with  mean ratios  derived from previous 
studies of BCDs (Thuan et al. \cite{til95}; Izotov \& Thuan \cite{IT99}).

\item The $^4$He mass fractions $Y$ = 0.246 $\pm$ 0.003 and 0.243 $\pm$ 0.010, 
derived for the brightest H {\sc ii} regions of SBS 1415+437, are in good 
agreement with previous determinations for this galaxy and with the 
primordial $^4$He mass fraction $Y_{\rm p}$ = (0.244 -- 0.245) $\pm$ 0.002 
(Izotov \& Thuan \cite{IT98}; Izotov et al. \cite{ICFGGT99}).
   
\item We use four different methods and different SF scenarios to derive
the age of the stellar populations in the LSB component of the galaxy.
For the outer LSB regions, 
the equivalent widths of the H$\alpha$ and H$\beta$ emission lines,
the equivalent widths of the H$\gamma$ and H$\delta$ absorption lines,
the spectral energy distributions and the ($V-I$) colours are 
reproduced quite well by models in which only a young stellar population
($t$ $\la$ 250 Myr) is present. For those regions, only
a small extinction with
$C$(H$\beta$) in the range 0 -- 0.1 is needed. An older stellar population,
if present,  does not contribute 
substantially to the optical luminosity of those outer regions.
For region {\it a}1, located closer to the brightest H {\sc ii} region,
a larger extinction ($C$(H$\beta$) = 0.22) is required to fit the 
observational data by the same young stellar population. 
The assumption of $C$(H$\beta$)=0 in all regions 
would require an increase of the relative fraction of the old stellar 
population from the outer LSB parts to the inner bright H {\sc ii}
regions, which would be contrary to the trend observed in the majority of 
BCDs, where the relative
contribution of old stars  increases outwards. 

Assuming no extinction, we find that the upper limit to the 
mass of the old
stellar population in SBS 1415+437, formed between 2.5 Gyr and 10 Gyr, is 
not greater than $\sim$ (1/20 -- 1) of the mass of the stellar population formed 
during the last $\sim$ 250 Myr. Depending on the region considered, this also 
implies that the star formation rate during the 
most recent star formation rate 
in SBS 1415+437 must be 20 to 1000 times greater than the SFR at ages $\ga$ 2.5 Gyr.

\item We compare the ($V-I$) colours of the LSB components and the oxygen abundances 
of  a sample of 26 low-metallicity dwarf irregular and BCD galaxies, 
including \sbs. It is shown 
that the LSB components are 
systematically bluer in the lower-metallicity galaxies. However, 
the observed trend is too steep to be explained only by metallicity effects.
Therefore, it is likely that lower-metallicity galaxies have also younger 
populations. In particular,
the luminosity-weighted ages of galaxies with LSB colours of ($V-I$) $\la$ 0.6 are probably 
not greater than 1--2 Gyr.
\end{enumerate}

\begin{acknowledgements}
N.G.G. has been supported by DFG grant 
436 UKR 17/2/02 and Y.I.I. acknowledges the G\"ottingen Academy of Sciences
for a Gauss professorship.
N.G.G. and Y.I.I. have been partially supported by  
Swiss SCOPE 7UKPJ62178 grant. They are grateful for the hospitality 
of the G\"ottingen Observatory. 
Y.I.I. and T.X.T. have been partially 
supported by NSF grant AST-02-05785.
 Research by P.P. and K.J.F. has been supported by the
Deutsches Zentrum f\"{u}r Luft-- und Raumfahrt e.V. (DLR) under
grant 50\ OR\  9907\ 7. 
 K.G.N. acknowledges the support from the Deutsche 
Forschungsgemeinschaft (DFG) grants FR 325/50-1 and FR 325/50-2.
This research has made use of the NASA/IPAC
Extragalactic Database (NED) which is operated by the Jet Propulsion
Laboratory, California Institute of Technology, under contract
with the National Aeronautics and Space Administration.
\end{acknowledgements} 

{}

\begin{thebibliography}{}
\bibitem[1984]{Aller84} Aller, L. H. 1984, Physics of Thermal Gaseous Nebulae,
   Dordrecht: Reidel
\bibitem[1989]{Anders89} Anders, E., \& Grevesse, N.
  1989, Geochim.Cosmochim.Acta, 53, 197
\bibitem[1990]{B90} Bessell, M. S. 1990, PASP, 102, 1181
\bibitem[1998]{B98} Bessell, M. S., Castelli, F., \& Plez, B. 1998, A\&A, 333, 231
\bibitem[1986]{Bica86} Bica, E., \& Alloin, D. 1986, A\&AS, 66, 171
\bibitem[1985]{Christ85} Christian, C. A., Adams, M., Barnes, J. V., et al. 1985, PASP, 97, 363
\bibitem[2002]{Crone02} Crone, M. M., Schulte-Ladbeck, R. E., Greggio, L., \& Hopp, U. 2002 ApJ, 567, 258
\bibitem[2001]{Dr01} Drozdovsky, I. O., Schulte-Ladbeck, R. E., Hopp, U., Crone, M. M., \& Greggio, L. 2001, ApJ, 551, L135
\bibitem[1998]{Esteban98} Esteban, C., Peimbert, M., Torres-Peimbert, S., \&
Escalante, V. 1998, MNRAS, 295, 401 
\bibitem[1999a]{Esteban99a} Esteban, C., Peimbert, M., Torres-Peimbert, S., 
Garcia-Rojas, J., \& Rodriguez, M. 1999a, ApJS, 120, 113
\bibitem[1999b]{Esteban99b} Esteban, C., Peimbert, M., Torres-Peimbert, S., \& 
Garcia-Rojas, J. 1999b, Rev. Mex. Astron. Astrofiz., 35, 65 
\bibitem[1997]{F97} Fioc, M., \& Rocca-Volmerange, B. 1997, A\&A, 326, 950
\bibitem[2001]{Fricke01} Fricke, K. J., Izotov, Y. I., Papaderos, P., 
Guseva, N. G., \& Thuan, T. X. 2001, AJ, 121, 169
\bibitem[1999]{GonLeith99b} Gonz\'alez Delgado, R. M., Leitherer, C., \&
 Heckman, T. M. 1999, ApJS, 125, 489
\bibitem[2001]{Guseva2001} Guseva, N. G., Izotov, Y. I., Papaderos, P., et al. 
2001, A\&A, 378, 756
\bibitem[2003a]{Guseva2003a} Guseva, N. G., Papaderos, P., Izotov, Y. I., et al. 2003a, A\&A, in press
\bibitem[2003b]{Guseva2003b} Guseva, N. G., Papaderos, P., Izotov, Y. I., et al. 2003b, A\&A, in press
\bibitem[1998]{IT98} Izotov, Y. I., \& Thuan, T. X. 1998, ApJ, 500, 188 
\bibitem[1999]{IT99} Izotov, Y. I., \& Thuan, T. X. 1999, ApJ, 511, 639
\bibitem[2002]{IT02} Izotov, Y. I., \& Thuan, T. X. 2002, ApJ, 567, 875 
\bibitem[1994]{ITL94} Izotov, Y. I., Thuan, T. X., \& Lipovetsky, V. A. 1994, 
 ApJ, 435, 647  
\bibitem[1997a]{ITL97a} Izotov, Y. I., Thuan, T. X., \& Lipovetsky, V. A. 1997a,
  ApJS, 108, 1
\bibitem[1997b]{ILCFGK97b} Izotov, Y. I., Lipovetsky, V. A., Chaffee, F. H., 
et al. 1997b, ApJ, 476, 698
\bibitem[1999]{ICFGGT99} Izotov, Y. I., Chaffee, F. H., Foltz, C. B., et al.
 1999, ApJ, 527, 757
\bibitem[2001a]{ICG01a} Izotov, Y. I., Chaffee, F. H., \& Green, R. F. 2001a, ApJ, 562, 727
\bibitem[2001b]{ICFTGPFG01b} Izotov, Y. I., Chaffee, F. H., Foltz, C. B., et al.
 2001b, ApJ, 560, 222
\bibitem[1996]{KobSkillman96} Kobulnicky, H. A., \& Skillman, E. D. 1996, ApJ, 471, 211
\bibitem[1999]{Kobul99} Kobulnicky, H. A., Kennicutt, R. C., Jr., 
\& Pizagno, J. L. 1999,  ApJ, 514, 544
\bibitem[2000]{Noeske00} Noeske, K. G., Guseva, N. G., Fricke, K. J., et al. 2000, A\&A, 361, 33
\bibitem[1996]{Papa96b} Papaderos, P., Loose, H.-H., Thuan, T. X., \& 
Fricke, K. J. 1996, A\&AS, 120, 207
\bibitem[1998]{Papa98} Papaderos, P., Izotov, Y. I., Fricke, K. J., Thuan, T. X., \& 
Guseva, N. G. 1998, A\&A, 338, 43
\bibitem[1999]{Papa99} Papaderos, P., Fricke, K. J., Thuan, T. X., Izotov, Y. I., \& Nicklas, H. 1999,  A\&A, 352, 57
\bibitem[2002]{papaderos02} Papaderos, P., Izotov, Y. I., Thuan, T. X., et al.
2002, A\&A, 393, 461 
\bibitem[2000]{pil00} Pilyugin, L. S. 2001, A\&A, 369, 594
\bibitem[1979]{Seaton79} Seaton, M. J. 1979, MNRAS, 187, 73 
\bibitem[1998]{Schlegel98} Schlegel, D. J., Finkbeiner, D. P., \& Davis, M. 1998, ApJ, 500, 525 
\bibitem[1998]{Schulte98} Schulte-Ladbeck, R. E.,  Crone, M. M., \& Hopp, U. 1998,
ApJ, 493, 23
\bibitem[1994]{Storchi94} Storchi-Bergmann, T., Calzetti, D., \& Kinney, A. L. 1994, ApJ, 429, 572
\bibitem[1995]{til95} Thuan, T. X., Izotov, Y. I., \& Lipovetsky, V. A. 1995, ApJ, 445, 108
\bibitem[1996]{til96} Thuan, T. X., Izotov, Y. I., \& Lipovetsky, V. A. 1996, ApJ, 463, 120
\bibitem[1999]{ti99} Thuan, T. X., Izotov, Y. I., \& Foltz, C. B. 1999, ApJ, 525, 105
\bibitem[1997]{Vaceli97} Vaceli, M. S., Viegas, S. M., Gruenwald, R., \& De Souza, R. E. 1997, AJ, 114, 1345
\bibitem[2003] {Sukyoung03} Yi, S. K. 2003, ApJ, 582, 202
\end{thebibliography}
\end{document}